\documentclass[11pt,a4paper]{article}
\pdfoutput=1
\usepackage{jheppub}
\usepackage{slashed}
\usepackage{comment}
\usepackage{color}
\usepackage{soul}

\makeatletter
\def\@fpheader{\relax}
\makeatother

\usepackage[czech,english]{babel}
\usepackage{graphicx}
\usepackage{amsmath,amsfonts,amssymb}
\DeclareMathOperator{\MyProd}{\scalebox{1.4}{$\mathrm{I\kern-0.2ex I}$}}

\title{Dark Solitons, $D$-branes and Noncommutative Tachyon Field Theory}

\author[a,\hspace{0.6mm} b]{Stefano Giaccari}

\emailAdd{sgiaccari@phy.hr}

\author[c,\hspace{0.6mm} d]{Jun Nian}

\emailAdd{nian@ihes.fr}

\affiliation[a]{Department of Physics, Faculty of Science\\
                    University of Zagreb\\
                    Bijeni\v{c}ka 32, HR-10000 Zagreb, Croatia\\}

\affiliation[b]{Department of Physics $\&$ Center for Field Theory and Particle Physics\\
                    Fudan University\\
                    200433 Shanghai, China\\}

\affiliation[c]{Institut des Hautes \'Etudes Scientifiques\\
	Le Bois-Marie, 35 route de Chartres\\
         91440 Bures-sur-Yvette, France\\}

\affiliation[d]{C.N. Yang Institute for Theoretical Physics\\
	Stony Brook University\\
         Stony Brook, NY 11794-3840, U.S.A.\\}

\abstract{In this paper we discuss the boson/vortex duality by mapping the (3+1)D Gross-Pitaevskii theory into an effective string theory in the presence of boundaries. Via the effective string theory, we find the Seiberg-Witten map between the commutative and the noncommutative tachyon field theories, and consequently identify their soliton solutions with $D$-branes in the effective string theory. We perform various checks of the duality map and the identification of soliton solutions. This new insight between the Gross-Pitaevskii theory and the effective string theory explains the similarity of these two systems at quantitative level.
}


\arxivnumber{}

\newcommand{\bea}{\begin{eqnarray}}
\newcommand{\eea}{\end{eqnarray}}

\newcommand{\be}{\begin{equation}}
\newcommand{\ee}{\end{equation}}

\begin{document}
\maketitle

\section{Introduction}\label{sec:introduction}

It has been known for a long time that some excitations in Bose-Einstein condensates (BEC), e.g. vortex lines, vortex rings and dark solitons, are very similar to some basic ingredients of string theory, such as open strings, closed strings and $D$-branes.

In Refs.~\cite{Nitta-1, Nitta-2}, the comparison between dark solitons and $D$-branes has been made quantitatively. The authors studied a two-component BEC model, and found the energy of this system is the same as the one in a four-dimensional $\mathcal{N}=2$ supersymmetric sigma model \cite{D-braneSigma}. Using the BPS procedure, they found the soliton solution and the vortex solution of the system. However, the soliton solution that they found in this configuration is not the standard dark soliton solution known in the Bose-Einstein condensates, instead it is the boundary between two components of the BEC.

More recently, some numerical work in BEC \cite{BEC} has demonstrated that the vortex lines can be attached directly to dark solitons, which mimics the configuration of open strings attached to $D$-branes in string theory. Hence, it is very conceivable that there should be some explanations about this similarity from theoretical point of view. If this correspondence can be put on solid ground, one may expect to simulate string theory in a BEC system, and at the same time bring in some new ideas to the study of Bose-Einstein condensates.

We would like to explore this relation between BEC and string theory from theoretical perspective. Our starting point is the following. The Gross-Pitaevskii equation is known as an effective theory to describe Bose-Einstein condensates \cite{PitaevskiiBook}, and it has been shown \cite{Zee, Franz, Gubser} that using the so-called boson/vortex duality for a spacetime without boundary the Gross-Pitaevskii equation can be mapped into an effective theory, which is very similar to the standard string theory in the large $B$-field limit. In this paper, we will demonstrate that this duality can also be generalized to a spacetime with boundary, where certain soliton solutions play the role of boundaries. By analyzing the effective string theory obtained from the duality, eventually we would like to identify dark soliton solutions to the Gross-Pitaevskii equation with $D$-branes in the effective string theory.

    \begin{figure}[!htb]
      \begin{center}
        \includegraphics[width=0.6\textwidth]{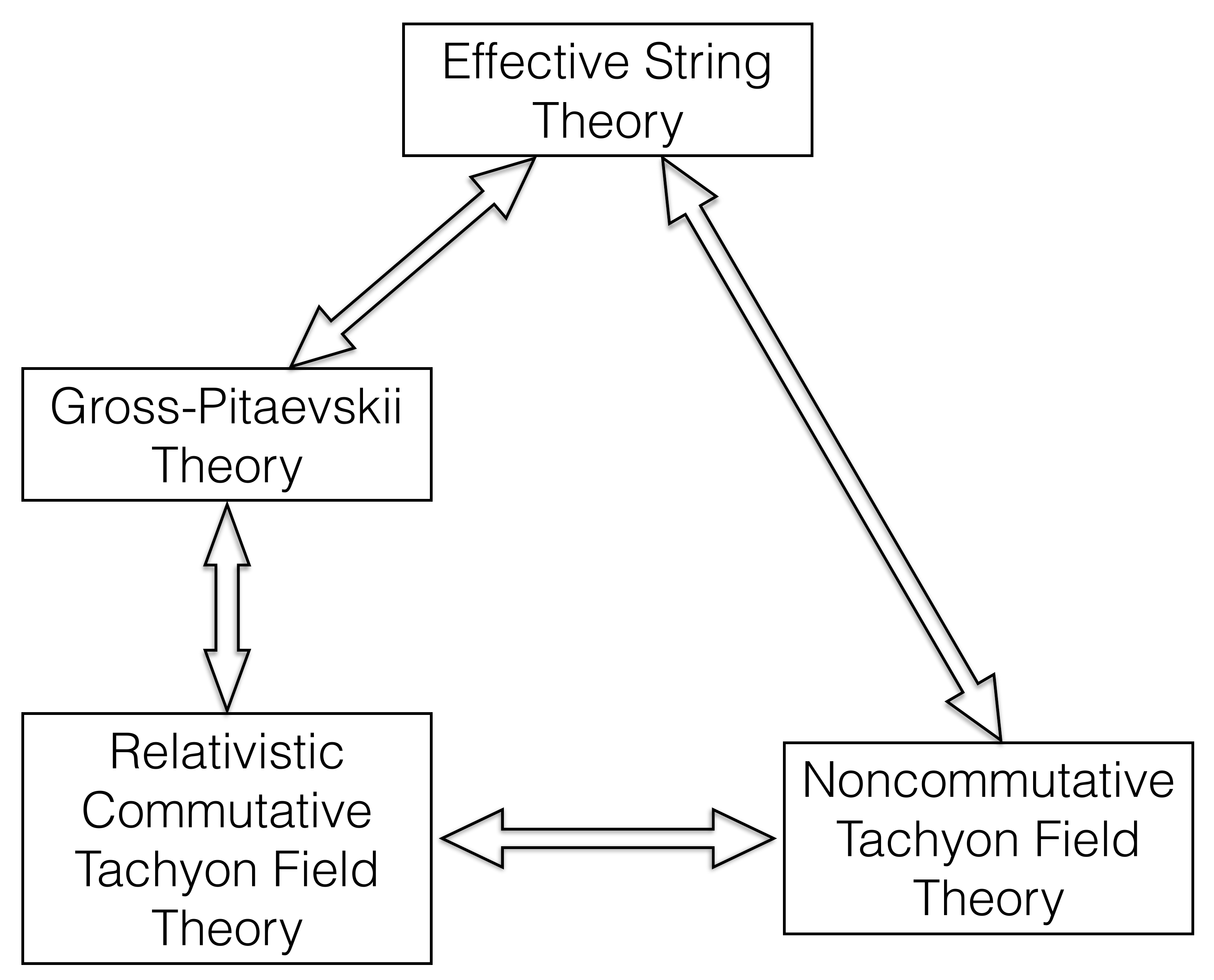}
      \caption{The relation between different theories}
      \label{fig:rel}
      \end{center}
    \end{figure}

Fig.~\ref{fig:rel} illustrates the relation among the theories and their soliton solutions, which will be discussed in this paper. As we discussed before, the Gross-Pitaevskii theory can be mapped into an effective string theory. It can also be viewed as the non-relativistic version of a relativistic commutative scalar field theory. Sometimes for convenience, we can gauge the relativistic commutative scalar field theory to obtain the commutative Abelian Higgs model. Although different in kinetic terms, the Gross-Pitaevskii theory and the commutative scalar field theory share the same time-independent dark soliton solution.

To identify dark soliton solutions with D-branes in the effective string theory, it is more convenient to first relate them to solitons in the noncommutative tachyon field theory, which has a natural relation with string theory. The approach of applying noncommutative geometry to string theory was first introduced by Seiberg and Witten \cite{SW}. They have studied the Yang-Mills theory from the open string sector in the large $B$-field limit, and have found that it admits both commutative and noncommutative descriptions, which are related by a nonlinear map of the corresponding gauge fields called the Seiberg-Witten map. A similar relation should be obtained for the tachyon field. If we think of the relativistic commutative scalar field theory as the commutative description of the tachyon field, there should be a noncommutative tachyon field theory obtained from the effective string theory. The commutative and the noncommutative tachyon fields can be related by a Seiberg-Witten map, and in this way the dark soliton solution to the commutative scalar field theory is related to the noncommutative soliton.

To find the Seiberg-Witten map for the tachyon field, one can first consider the commutative and the noncommutative Abelian Higgs models. The commutative Abelian Higgs model has been studied in the literature in great detail. It is known that this theory has some topologically nontrivial solutions such as the Nielsen-Olesen vortex line \cite{NOvortex}, which becomes the vortex line solution in the Gross-Pitaevskii theory when the gauge field is turned off, and the endpoints of the Nielson-Olsen string have to terminate at (anti-)monopoles \cite{DanishBook}. On the other hand, the (2+1)-dimensional noncommutative Abelian Higgs model and its vortex solutions have been studied in Ref.~\cite{AHM}. It is straightforward to generalize the analysis in Ref.~\cite{AHM} to our case and find the Seiberg-Witten map between the tachyons. We then turn off the gauge fields to obtain the Seiberg-Witten map for pure scalar field theories, which are just the commutative and the noncommutative tachyon field theories that we are looking for. It can be shown that the noncommutative scalar field theory obtained in this way coincides with the noncommutative tachyon field theory obtained from the effective string theory at the leading order. For this theory, Ref.~\cite{GMS} has studied its soliton solutions, and then Ref.~\cite{Harvey} has shown that these noncommutative solitons can be identified with the $D$-branes in string theoy.

Hence, after all these steps above the circle shown Fig.~\ref{fig:rel} is closed, and under certain approximations we can identify dark solitons in the Gross-Pitaevskii theory with $D$-branes in the effective string theory as well as solitons in the noncommutative tachyon field theory.

This paper is organized as follows. In Section~\ref{sec:scalar}, we review the commutative Abelian Higgs model, the Gross-Pitaevskii theory and some solutions with nontrivial topology. In Section~\ref{sec:duality}, we map the Gross-Pitaevskii theory into an effective string theory in the presence of boundaries. As discussed above, we would like to identify dark solitons, noncommutative solitons and $D$-branes. To test this identification, we compare different descriptions and their solitons in Section~\ref{sec:soliton}. The noncommutative Abelian Higgs model is discussed in Subsection~\ref{NcAHM}. By turning off the gauge field, we obtain a noncommutative scalar field theory, which can be interpreted as a noncommutative tachyon field theory. In Subsection~\ref{noncommTFT}, we obtain the noncommutative tachyon field theory directly from the effective string theory. The tensions of solitons are compared with the ones of $D$-branes from different descriptions in Subsection~\ref{NcAHM} and \ref{noncommTFT}. Moreover, the comparison of the $D$-brane interaction with the numerical results of the dark soliton interaction is made in Subsection~\ref{check}. Finally, future directions are discussed in Section~\ref{sec:discussion}. Some details of the computation will be presented in a few appendices. In Appendix~\ref{app:sol} we list some solutions with nontrivial topology to the Gross-Pitaevskii theory. The derivations of the effective string theory with boundaries from the Gross-Pitaevskii theory are presented in Appendix~\ref{app:duality}. The tachyon potential in the presence of the $B$-field will be discussed in Appendix~\ref{app:TachyonPot}.

\section{Commutative Field Theories}\label{sec:scalar}

In this section, we would like to review some known results of commutative scalar field theories. We first discuss the commutative Abelian Higgs model and the Nielsen-Olsen vortex line solution in Subsection~\ref{commAHM}. By turning off the gauge field, we obtain the relativistic commutative tachyon field theory, whose non-relativistic version is the Gross-Pitaevskii theory. We will discuss this theory and some solutions with nontrivial topology in Subsection~\ref{commTFT}.

\subsection{Abelian Higgs Model}\label{commAHM}
The commutative Abelian Higgs model in (3+1)-dimensions is given by the Lagrangian:
\be\label{eq:commAHM}
  \mathcal{L} = -\frac{1}{4} F_{\mu\nu} F^{\mu\nu} - 
  \frac{1}{2}
   |(\partial_\mu + i e A_\mu) \phi|^2 - \lambda (|\phi|^2 - |\phi_0|^2)^2\, ,
\ee
where $\phi$ is a complex scalar, and $A_\mu$ is the gauge field. The equations of motion are
\be
  (\partial_\mu + i e A_\mu)^2 \phi - 4 \lambda (|\phi|^2 - |\phi_0|^2) \phi = 0\, ,
    \label{EOMscalar}
\ee
\be
  \partial^\nu F_{\mu\nu} = \frac{i}2 e \left(\phi^* \partial_\mu \phi - \phi \partial_\mu \phi^* \right) - e^2 A_\mu \phi^* \phi\, .
\ee

This theory has nontrivial classical solutions, including string-like Nielsen-Olsen vortex lines. The Nielsen-Olsen vortex line solution (or Nielsen-Olsen string) was found first by Nielsen and Olsen in Ref.~\cite{NOvortex}. Under the gauge $A_0 = 0$, considering a string-like solution, i.e. preserving a cynlindrical symmetry, one can assume the axis along the $z$-direction and apply the ansatz
\be\label{eq:NOansatz}
  \bold{A} (\bold{r}) = \frac{\bold{r} \times \bold{e}_z}{r} |\bold{A} (r)|\, ,
\ee
and the flux carried by the Nielsen-Olsen vortex line is
\be
  \Phi (r) = \oint A_\mu (x)\, d x^\mu = 2 \pi r |\bold{A} (r)|\, ,
\ee
Since the covariant derivative on the scalar field vanishes outside the vortex line, the phase of $\phi$ defined by $\phi = |\phi| \, e^{i \eta}$ satisfies
\be
  d \eta + e A = 0 \quad \textrm{ for } r \to \infty\, .
  \label{IReom}
\ee
Hence, the property that $\phi$ should be single-valued requires
\be
  \Phi = \lim_{r_0 \to \infty} \oint_{r = r_0} A = - \lim_{r_0 \to \infty} \frac{1}{e} \oint_{r = r_0} d \eta = n \frac{2 \pi}{e}\, ,
  \label{MagFlux}
\ee
i.e. the magnetic flux carried by the Nielsen-Olsen vortex line is quantized.

By plugging the ansatz \eqref{eq:NOansatz} into the Lagrangian \eqref{eq:commAHM} and perfomring the variation of the fields, one can obtain the classical equations of motion for the configurations with cynlindrical symmetry, which can be solved numerically. The solutions have the asymptotic behavior:
\be
  |\phi| = |\phi_0| = \textrm{const}\, ,\quad |\bold{A}| = \frac{1}{er} + \frac{c}{e}\, K_1 (e r |\phi|)\, , \textrm{ for } r \to \infty\, ;
\ee
\be
  |\phi| = 0\, ,\quad \textrm{ for } r \to 0\, .
\ee

\begin{figure}[!htb]
      \begin{center}
        \includegraphics[width=0.33\textwidth]{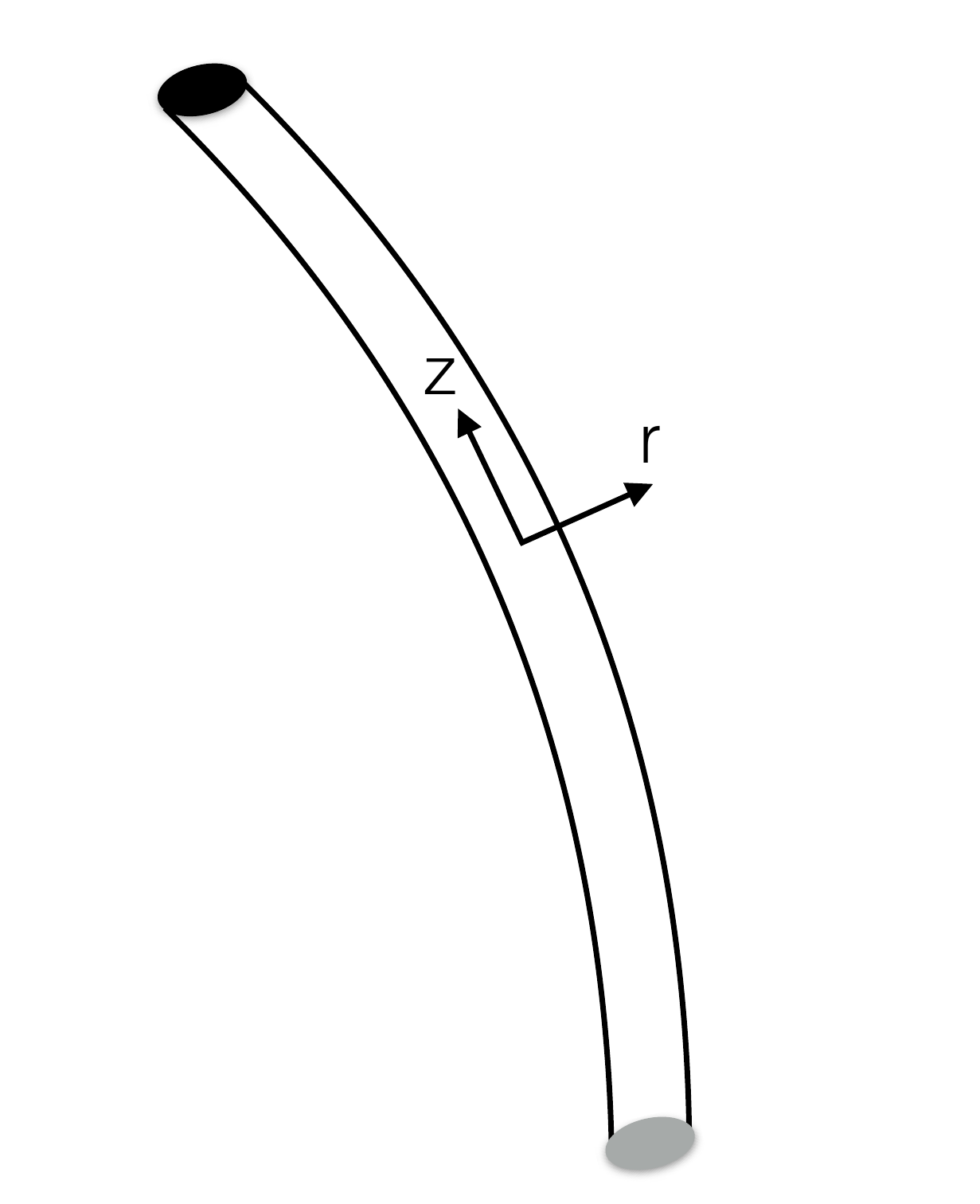}
      \caption{The sketch of the Nielsen-Olsen vortex line ending on a monopole anti-monopole pair}
      \label{fig:NOvortex}
      \end{center}
    \end{figure}

As discussed in Ref.~\cite{DanishBook}, an infinitely long vortex line has infinite energy, which is unphysical. In order to have a finite length, a vortex line can terminate at a magnetic (anti-)monopole, which has the magnetic charge $g = n\, 2 \pi / e$. Hence, the magnetic flux carried by the Nielsen-Olsen vortex line can be absorbed by the monopole anti-monopole pair. Moreover, one can demonstrate that in this case the potential between the monopole anti-monopole pair is linear in the distance between them, i.e., the Nielsen-Olsen vortex line realizes the confinement in the Abelian Higgs model. Schematically, the configuration of a finite Nielsen-Olsen vortex line with a monopole anti-monopole pair at two endpoints is shown in Fig.~\ref{fig:NOvortex}.

It is also known that the Nielsen-Olsen vortex line can be effectively described by the Nambu string action \cite{NOvortex, DanishBook}, which also confirms our following discussions that the scalar theory can be mapped into an effective string theory. As we will also see later, the vortex line solution to the Gross-Pitaevskii equation can be viewed as the Nielsen-Olsen vortex line solution when the gauge field is turned off.

\subsection{Commutative Scalar Field Theory}\label{commTFT}

There are two kinds of commutative scalar field theory, relativistic and non-relativistic. Both of them will appear in the rest of this paper. If some solutions are time-independent, they are solutions to both the relativistic and the non-relativistic commutative scalar field theory.

Let us first look at the relativistic one. If we turn off the gauge field, the commutative Abelian Higgs model \eqref{eq:commAHM} becomes the corresponding relativistic commutative scalar field theory:
\be\label{eq:LcommScalar}
  \mathcal{L}_\textrm{c} = -\frac12 |\partial_\mu \phi|^2 -\lambda (|\phi|^2 - |\phi_0|^2)^2\, ,
\ee
where the subscript ``c'' stands for the commutative theory, in contrast to the noncommutative theory that we will discuss later.

The commutative scalar theory \eqref{eq:LcommScalar} can be viewed as the relativistic version of the Gross-Pitaevskii (GP) theory, which is given by the following non-relativistic Lagrangian:
\be\label{eq:LGP}
  \mathcal{L}_{GP} = i \phi^\dagger \partial_t \phi - \frac{1}{2m} (\nabla \phi^\dagger) (\nabla \phi) - \frac{g}{2} (|\phi|^2 - \rho_0)^2\, .
\ee
Varying it with respect to $\phi^\dagger$, we obtain the Gross-Pitaevskii equation:
\be\label{eq:GPeq}
  i\, \partial_t \phi + \frac{1}{2 m} \nabla^2 \phi - g\, (|\phi|^2 - \rho_0)\, \phi = 0\, .
\ee
The Gross-Pitaevskii equation has various solutions with nontrivial topology \cite{PitaevskiiBook}. Let us briefly summarize them in the following, and more details of these solutions can be found in Appendix~\ref{app:sol}.
\begin{itemize}
\item Dark soliton:

For the repulsive interaction, i.e. $g>0$ in Eq.~\eqref{eq:GPeq}, there is a time-independent solution, which can be characterized by a static plane with two spatial directions, and for the direction $x$ normal to the plane it has the following profile:
\be\label{eq:darksoliton}
  \phi (x) = \frac{\hbar \sqrt{n}}{\sqrt{m}} \, \textrm{tanh}\, \left[\frac{x}{\sqrt{2} \xi} \right]\, ,
\ee
where $n$ is the density at infinity, and $\xi \equiv \hbar / \sqrt{2 m g n}$ is a parameter called healing length. At the position $x=0$ where the dark soliton is localized, the density is zero. Consequently, in cold atom experiments that simulate the Gross-Pitaevskii equation, this kind of solution appears dark. That is why it is called ``dark soliton''.

\item Grey soliton:

The grey soliton solution is similar to the dark soliton solution, but is a time-dependent moving plane extended in two spatial directions. From the dark soliton solution discussed above, one can easily perform a Galilean boost to obtain a moving grey soliton solution given by
\be
  \Psi (x - v t) = \frac{\hbar \sqrt{n}}{\sqrt{m}} \left(i \frac{v}{c} + \sqrt{1 - \frac{v^2}{c^2}}\, \textrm{tanh} \left[\frac{x - v t}{\sqrt{2} \xi} \, \sqrt{1 - \frac{v^2}{c^2}} \right] \right)\, ,
\ee
where $c$ and $v$ stand for the speed of sound and the speed of the grey soliton respectively. At the center $x=0$ the density is non-zero, which makes it appear grey in real experiments. That is where its name comes from.

\item Bright soliton:
For the attractive interaction, i.e. $g<0$ in Eq.~\eqref{eq:GPeq}, there is a different kind of soliton solution called bright soliton, which is given by
\be
  \phi(x) = \phi(0) \frac{1}{\textrm{cosh} (x / \sqrt{2} \xi)}\, .
\ee
In contrast to the dark and the grey soliton, the bright soliton solution has the maximal density at the center, which appears bright in real experiments.

\item Vortex line:
Unlike the various plane-like solutions discussed above, the vortex line is a string-like solution, which in the cylindrical coordinates $(\eta,\, \varphi)$ is given by
\be
  \phi = \sqrt{n} \, f(\eta) \, e^{i s \varphi}\, ,
  \label{eq:VortexLine}
\ee
where $n$ is the density, and $f(\eta)$ satisfies the following equation:
\be
  \frac{1}{\eta} \frac{d}{d\eta} \left(\eta \frac{df}{d\eta} \right) + \left(1 - \frac{s^2}{\eta^2} \right) f - f^3 = 0\, .
\ee
For given boundary conditions and a fixed value of $s$, the equation above can be solved numerically.

\item Vortex ring:
The vortex line solution has two endpoints. These two end points can join together, and the original vortex line solution becomes a ring-like object, which is called vortex ring. Besides different shapes, it turns out that the vortex ring cannot be at rest in contrast to the vortex line solution. The collision of two vortex rings has been studied in Ref.~\cite{Gubser} using the boson/vortex duality.

\end{itemize}

We would like to emphasize that, although the Gross-Pitaevskii equation \eqref{eq:GPeq} is a non-relativistic version of the field equation for the commutative scalar field theory \eqref{eq:LcommScalar}, they share the same time-independent dark soliton solution \eqref{eq:darksoliton} discussed above.

According to the well-known Derrick's theorem, stable topologically nontrivial nonsingular field configurations for a purely scalar field theory with second derivatives can exist only for $d<2$, where $d$ is the spatial dimension of the configuration. Hence, the dark soliton solution is an unstable configuration. In practice, one can restrict the size of transverse dimensions of the dark soliton to make it stable.

In principle, the solutions with nontrivial topology discussed in this section can be classified using the homotopy group, which is beyond the scope of this paper but has been studied in great detail in the literautre.

\section{Boson/Vortex Duality with Boundaries}\label{sec:duality}

As we mentioned before, it has been found recently in Ref.~\cite{BEC} that vortex lines can be attached to the dark soliton planes to form a relatively stable configuration. If we treat the vortex line solution as an effective string, we should be able to describe this novel stable configuration using an effective string theory, where dark solitons play the role of $D$-branes. In this section, we briefly review the mapping of the Gross-Pitaevskii equation into an effective string theory in the presence of dark solitons as boundaries. The identification of dark solitons with $D$-branes will be justified in later sections.

It has been shown in Refs.~\cite{Zee, Franz, Gubser} that without boundaries the (3+1)D Gross-Pitaevskii theory \eqref{eq:LGP} can be mapped into an effective string theory, and similar analysis can be generalized to other dimensions (see e.g. Ref.~\cite{Beekman:2010zx}). In the presence of boundaries, some subtleties and new features emerge and have to be paid special attention to.

In this paper we consider the following simplest configuration. Two parallel dark solitons are placed in the bulk of the (3+1)D spacetime, which are separated in the spatial $z$-direction with a distance $L$. The dark soliton plane can be viewed as a (2+1)D spacetime without boundary. In the space between two dark soliton planes there can be closed vortex lines and open vortex lines, and the endpoints of the open vortex lines have to be attached to one of the dark soliton planes.

Now let us discuss the duality map in the presence of dark solitons, which is also discussed in Ref.~\cite{BEC} using an alternative approach. For the configuration with boundaries, we can repeat the first few steps for the configuration without boundaries discussed in Ref.~\cite{Gubser}. Let us recall the Gross-Pitaevskii Lagrangian \eqref{eq:LGP}:
\begin{displaymath}
  \mathcal{L}_{GP} = i \phi^\dagger \partial_t \phi - \frac{1}{2m} (\nabla \phi^\dagger) (\nabla \phi) - \frac{g}{2} (|\phi|^2 - \rho_0)^2\, .
\end{displaymath}
With the parametrization:
\be\label{eq:Param}
  \phi = \sqrt{\rho}\, e^{i \eta}\, ,
\ee
the original Gross-Pitaevskii Lagrangian \eqref{eq:LGP} becomes
\be\label{eq:LGPrhoeta}
  \mathcal{L} = \frac{i \dot{\rho}}{2} - \rho \dot{\eta} - \frac{\rho}{2m} (\nabla \eta)^2 - \frac{(\nabla \rho)^2}{8 m \rho} - \frac{g}{2} (\rho - \rho_0)^2\, .
\ee
We can drop the first term as a total derivative and separate the phase-dependent part from the phase-independent part:
\begin{align}
  \mathcal{L}_1 & \equiv - \rho \dot{\eta} - \frac{\rho}{2m} (\nabla \eta)^2\, ,\\
  \mathcal{L}_2 & \equiv - \frac{(\nabla \rho)^2}{8 m \rho} - \frac{g}{2} (\rho - \rho_0)^2\, .
\end{align}

In the presence of a boundary, the phase $\eta$ consists of two parts, one part $\eta (t, x, y, z)$ defined on the (3+1)D spacetime and the other part $\eta (t, x, y)$ defined only on the (2+1)D boundary, which can be the dark soliton plane. Both of them may contain singularities. We can perform the duality map on the (3+1)D and the (2+1)D spacetime separately, i.e.,
\be\label{eq:3D4DAction}
  S = \int d^4 x \left[\mathcal{L}_1^{4D} + \mathcal{L}_2^{4D} \right] + \ell \int d^3 x \left[\mathcal{L}_1^{3D} + \mathcal{L}_2^{3D} \right]\, ,
\ee
where $\ell$ is a constant length scale due to the dimensional reason, which will be discussed later in this subsection.

The separation of the (3+1)D part and the (2+1)D part of the action can also be obtained in the following way. Instead of Eq.~\eqref{eq:Param} let us use another parametrization:
\be\label{eq:ParamNew}
  \phi = p \, e^{i \eta}\, ,
\ee
where $p = \sqrt{\rho}$. For a dark soliton given by Eq.~\eqref{eq:darksoliton}, the background part of the factor $p$ only depends on $z$:
\be\label{eq:fDS}
  p_0 = \sqrt{n}\, \textrm{tanh} \left(\frac{z}{\sqrt{2} \ell} \right)\, .
\ee
It is usually said that there is a $\pi$-jump in the phase when across a dark soliton plane. This is just due to the fact that $p$ changes sign from one side of the soliton to the other side, and one can absorb the sign into the phase using $-1 = \textrm{exp} (i \pi)$. Equivalently, we can keep the sign change of $p$ and consider the phase without a $\pi$-jump. Using the new parametrization \eqref{eq:ParamNew}, we can rewrite the GP Lagrangian \eqref{eq:LGP} as
\be\label{eq:LGPfeta}
  \mathcal{L} = i p \dot{p} - p^2 \dot{\eta} - \frac{p^2}{2m} (\nabla \eta)^2 - \frac{(\nabla p)^2}{2m} - \frac{g}{2} (p^2 - p_c^2)^2\, ,
\ee
which can also be obtained by replacing $\rho$ with $p^2$ in Eq.~\eqref{eq:LGPrhoeta}, and $p_c = \sqrt{n}$ is a constant. Again, the first term is a total derivative that can be dropped. In Appendix~\ref{app:duality}, we analyze Eq.~\eqref{eq:LGPfeta} term by term under the assumption that $p$ consists of both the background and the fluctuations, i.e.,
  \be
    p = p_0 + \widetilde{p}\, ,
  \ee
where we take $p_0$ to be the profile given by Eq.~\eqref{eq:fDS}, which depends only on $z$, while we assume that $\widetilde{p}$ does not depend on $z$. The physical reason is that the dark soliton is very heavy, so that its longitudinal position is fixed, and there are no fluctuations in the longitudinal direction. As shown in Appendix~\ref{app:duality}, in the background of a dark soliton, the action can be expressed as
\be
  \int d^4 x \, \mathcal{L} = \ell \int d^3 x\, \left[- \widetilde{\rho} \dot{\eta} - \frac{1}{2m} \left(\widetilde{\rho} + C \right)(\widetilde{\nabla} \eta)^2 - \frac{1}{8m \widetilde{\rho}} \left( \widetilde{\nabla} \widetilde{\rho} \right)^2 - \frac{g}{2} \left(\widetilde{\rho} - \widetilde{\rho}_0 \right)^2 \right]\, ,
  \label{3dAction}
\ee
where $C$ is a constant. This expression corresponds to the 3D part of the action, and justifies the separation presented in Eq.~\eqref{eq:3D4DAction}.

As shown in Eq.~\eqref{eq:3D4DAction}, in the presence of boundaries the full theory should include both the (3+1)D duality and the (2+1)D duality. In the bulk one can still perform the (3+1)D duality is exactly the same as the case without boundaries discussed in Ref.~\cite{Gubser}. Hence, we focus on the (2+1)D duality map in the following. Note that all the fields with tilde ($\widetilde{\phantom{a}}$) are defined in (2+1)D, i.e., they depend only on $(t,\, x,\, y)$ but are independent of the longitudinal coordinate $z$.

The steps are similar to the (3+1)D case. One can introduce a 3-vector $f^a = (\rho,\, f^{\hat{a}})$ with $a \in \{t, x, y \}$ and $\hat{a} \in \{x,\, y \}$. Assuming that
\be
  \int \mathcal{D} f^{\hat{a}}\, \textrm{exp} \left(i \ell \int d^3 x\, \frac{m}{2 \widetilde{\rho}\,' } f^{\hat{a}} f_{\hat{a}} \right) = 1\, ,
\ee
we can rewrite the first two terms in the action \eqref{3dAction} without changing the path integral in the following way:
\be
  - \widetilde{\rho} \dot{\eta} - \frac{\widetilde{\rho}\,'}{2m} (\widetilde{\nabla} \eta)^2 + \frac{m}{2\widetilde{\rho}\,'} \left(f_{\hat{a}} - \frac{\widetilde{\rho}\,'}{m} \widetilde{\nabla}_{\hat{a}} \eta \right)^2 = - \widetilde{\rho} \dot{\eta} + \frac{m}{2 \widetilde{\rho}\,' } f^{\hat{a}} f_{\hat{a}} - f^{\hat{a}} \partial_{\hat{a}} \eta = \frac{m}{2 \widetilde{\rho}\,' } f^{\hat{a}} f_{\hat{a}} - f^a \partial_a \eta\, ,
\ee
where $\widetilde{\rho}' \equiv \widetilde{\rho} + C$.

Now one can separate the (2+1)D phase $\eta (t,\, x,\, y)$ into the smooth part and the singular part:
\be\label{eq:3Detaterm}
  - f^a \partial_a \eta = - f^a \partial_a \eta_{\textrm{smooth}} - f^a \partial_a \eta_{\textrm{singular}}\, .
\ee
The smooth part $\eta_{\textrm{smooth}}$ does not feel the (2+1)D vortices from the endpoints of vortex lines, and it is well-defined on the whole dark soliton plane. Integrating it out, we obtain
\be
  \partial_a f^a = 0\, ,
\ee
which can be solved by
\be
  f^a = \frac{1}{2} \epsilon^{abc} F_{bc}\quad \textrm{with} \quad F_{ab} = \frac{1}{2} (\partial_a A_b - \partial_b A_a)\, .
\ee
Consequently, the contribution of $\eta_{\textrm{smooth}}$ to the first two terms in the action \eqref{3dAction} is
\begin{align}
  \textrm{exp} \left[i \int d^3 x\, \mathcal{L}_1^{3D} \right] & = \int \mathcal{D} f^a \, \textrm{exp} \Bigg\{i\ell \int d^3 x\, \left[- f^a \partial_a \eta + \frac{m}{2 \widetilde{\rho}} f^{\hat{a}} f_{\hat{a}} \right] \Bigg\} \nonumber\\
  {} & = \int \mathcal{D} f^a \, \textrm{exp} \Bigg\{i\ell \int d^3 x\, \left[- f^a \partial_a \eta_{\textrm{singular}} + \frac{m}{2 \widetilde{\rho}} F_{0 \hat{a}}^2 \right] \Bigg\} \nonumber\\
  {} & = \int \mathcal{D} f^a \, \textrm{exp} \Bigg\{i\ell \int d^3 x\, \left[- f^a \partial_a \eta_{\textrm{singular}} + \frac{m}{2 \widetilde{\rho}} \widetilde{F}_{0 \hat{a}}^2 \right] \Bigg\} \, ,
\end{align}
and the remaining terms in the action \eqref{3dAction} are
\be
  \mathcal{L}_2^{3D} = - \frac{(\nabla \widetilde{F}_{\hat{a} \hat{b}})^2}{16 m \widetilde{\rho}} - \frac{g}{4} \widetilde{F}_{\hat{a} \hat{b}}^2\, ,
\ee
where $\hat{a},\, \hat{b} \in \{x, y \}$, and $\widetilde{F}$ denotes the fluctuation of the field strength $F$. Together, they form
\begin{align}
  {} & \int \mathcal{D} \widetilde{\rho} \mathcal{D} \eta \, \textrm{exp} \Bigg\{i \ell \int d^3 x \left[- \widetilde{\rho} \dot{\eta} - \frac{\widetilde{\rho}}{2m} (\nabla \eta)^2 - \frac{(\nabla \widetilde{\rho})^2}{8 m \widetilde{\rho}} - \frac{g}{2} (\widetilde{\rho} - \rho_0)^2 \right] \Bigg\} \nonumber\\
  = & \, \int \mathcal{D} A_a \, \textrm{exp} \Bigg\{i \ell \int d^3 x \left[- f^a \partial_a \eta_{\textrm{singular}} - \frac{g}{4} \eta^{ab} \eta^{cd} \widetilde{F}_{ac} \widetilde{F}_{bd} - \frac{(\nabla \widetilde{F}_{\hat{a} \hat{b}})^2}{16 m \widetilde{\rho}} \right] \Bigg\}\, ,
\end{align}
where $\eta_{ab} = (- \widetilde{\rho} g / (2m),\, 1,\, 1)$ is an effective 3D metric.

Now let us consider the term $- f^a \partial_a \eta_{\textrm{singular}}$. Since we have found the solution $f^a = \epsilon^{abc} \partial_b A_c$, it can be plugged into the singular term, then we obtain
\begin{align}
  {} & i \ell \int d^3 x\,\left[- f^a \partial_a \eta_{\textrm{singular}} \right] \nonumber\\
  = & \, -i \ell \int d^3 x\, \epsilon^{abc} (\partial_b A_c) \partial_a \eta_{\textrm{singular}} \nonumber\\
  = & \, - i \ell \int d^3 x\, \epsilon^{cab} A_c (\partial_a \partial_b \eta_{\textrm{singular}}) \nonumber\\
  = & \, - 2\pi i \ell \int d^3 x\, A_c j^c\, ,
\end{align}
where we have defined a vortex current:
\be
  j^c \equiv \frac{1}{2 \pi} \epsilon^{cab} \partial_a \partial_b \eta_{\textrm{singular}}\, ,
\ee
which satisfies
\be
  \int d^2 x \, j^0 = \frac{1}{2\pi} \int d^2 x\, \epsilon^{\hat{a} \hat{b}} \partial_{\hat{a}} \partial_{\hat{b}} \eta_{\textrm{singular}} = \frac{1}{2\pi} \int d^2 x\, \nabla \times (\nabla \eta_{\textrm{singular}}) = \frac{1}{2\pi} \oint d\vec{x} \cdot \nabla \eta_{\textrm{singular}} = 1\, .
\ee
Because the vortices on the dark soliton plane can also be viewed as the endpoints of the vortex lines in the (3+1)D spacetime, we may also use the following relation in (3+1)D:
\be
  \epsilon^{\lambda\sigma\mu\nu} \partial_\mu \partial_\nu \eta = - 2\pi \int d^2 \sigma\, \epsilon^{\alpha\beta} \partial_\alpha X^\lambda \partial_\beta X^\sigma \, \delta^4 (x^\mu - X^\mu) 
\ee
to obtain
\begin{align}
  {} & i \ell \int d^3 x\,\left[- f^a \partial_a \eta_{\textrm{singular}} \right] \nonumber\\
  = & \, - i \ell \int d^3 x\, \epsilon^{zcab} A_c (\partial_a \partial_b \eta_{\textrm{singular}}) \nonumber\\
  = & \, 2 \pi i \ell \int d^3 x\, \int d\tau d\sigma\, A_a \epsilon^{\alpha\beta} \partial_\alpha X^z \partial_\beta X^a \, \frac{1}{L} \delta^3 (x - X) \nonumber\\
  = & \, 2 \pi i \ell \int d\tau\, A_a \partial_\tau X^a\, ,
\end{align}
where the 3D $\delta$-function is related to the 4D $\delta$-function in the following way:
\be
  \delta^4 (x - X) = \frac{1}{L} \delta^3 (x - X)\, .
\ee
In the last step we have used the fact that only $\frac{\partial X^z}{\partial \sigma} = \frac{\partial z}{\partial \sigma}$ is nonvanishing, when the vortex line is perpendicular to the dark soliton, i.e. $z \parallel \sigma$, hence,
\be
  \int d\sigma \partial_\sigma X^z = \int d\sigma \frac{\partial z}{\partial \sigma} = \int dz = L\, ,
\ee
where $L$ is the distance between two parallel dark soliton planes.

There are still two issues that we have to address carefully. One is the dimensionality. In the following we list the mass dimensions of various fields and parameters:
\be
  [\rho] = 3\, ,\quad [\eta] = 0\, ,\quad [m] = 1\, ,\quad [g] = -2\, ,
\ee
\be
  [H] = 3\, ,\quad [B] = 2\, ,\quad [\widetilde{F}] = 3\, ,\quad [A] = 2\, .
\ee
Conventionally, the gauge field $A$ has mass dimension 1, which can be achieved by absorbing the length scale $\ell$ into $A$, i.e.,
\be
  \ell A_a \rightarrow A_a\, ,
\ee
to make it of dimension 1. Also, the 2-form gauge field $B_{\mu\nu}$ is conventionally dimensionless. To achieve it, we can separate a dimensionful constant from it, i.e.,
\be
  B_{\mu\nu} \rightarrow \frac{1}{2\pi \alpha'} B_{\mu\nu}\, ,
\ee
where $\alpha' = \ell_s^2$ with $\ell_s$ denoting the effective string length scale.

The other issue is that we treat the (3+1)D duality and the (2+1)D duality separately. Although they seem to be independent of each other, the dualities are related through a boundary integral in (3+1)D. Let us recall for the (3+1)D case:
\be\label{eq:bdyterms}
  \mathcal{L}^{4D} \supset -f^\mu \partial_\mu \eta_{\textrm{smooth}} - f^\mu \partial_\mu \eta_{\textrm{vortex}}\, .
\ee
We can partially integrate the first term and then integrate out $\eta_{\textrm{smooth}}$ to obtain the equation $\partial_\mu f^\mu = 0$. During the derivation we drop a boundary term
\be
  -\int d^4 x\, \partial_\mu (f^\mu \eta_{\textrm{smooth}})\, ,
\ee
which vanishes when the dark soliton is absent. In the presence of the dark soliton, this boundary term becomes
\begin{align}
  -\int d^4 x\, \partial_\mu (f^\mu \eta_{\textrm{smooth}}) & = -\int d^3 x\, f^\mu \eta_{\textrm{smooth}} \nonumber\\
  {} & = -\int d^3 x\, \epsilon^{zabc} \frac{1}{2} (\partial_a B_{bc}) \eta_{\textrm{smooth}} \nonumber\\
  {} & = \int d^3 x\, \epsilon^{abc} \frac{1}{2} B_{bc} \partial_a \eta_{\textrm{smooth}}\, ,
\end{align}
which should be combined with the term $-f^a \partial_a \eta_{\textrm{smooth}}$ in the (2+1)D duality. Hence, precisely speaking, for the (2+1)D duality instead of Eq.~\eqref{eq:3Detaterm} there should be
\be
  \mathcal{L}^{3D} \supset - \left(f^a - \frac{1}{2} \epsilon^{abc} B_{bc} \right) \partial_a \eta_{\textrm{smooth}} - f^a \partial_a \eta_{\textrm{singular}}\, .
\ee
Partially integrating the first term, instead of $\partial_a f^a = 0$ we will obtain
\be
  \partial_a \left(f^a - \frac{1}{2} \epsilon^{abc} B_{bc} \right) = 0\, .
\ee
The solution to this equation is
\be
  f^a - \frac{1}{2} \epsilon^{abc} B_{bc} = \frac{1}{2} \epsilon^{abc} F_{bc}\quad \textrm{with}\quad F_{bc} = \frac{1}{2} (\partial_b A_c - \partial_c A_b)\, .
\ee
Consequently,
\be
  f^a = \frac{1}{2} \epsilon^{abc} (F_{bc} + B_{bc})\, .
\ee
Hence, in the (2+1)D duality that we discussed above $F$ should be replaced by $F+B$. After introducing some length scales to match the conventional dimensions, the combination should be
\be
  F_{ab} + \frac{1}{2 \pi \alpha'} B_{ab}\, .
\ee

Therefore, the final expression of the (2+1)D dual theory is
\begin{align}
  {} & \int \mathcal{D} A_a \, \textrm{exp} \Bigg\{- 2 \pi i \int d^3 x\, A_a j^a + \frac{i}{\ell} \int d^3 x \left[- \frac{g}{4} (\widetilde{F} + \frac{1}{2 \pi \alpha'} \widetilde{B})^2 - \frac{(\nabla \widetilde{F}_{\hat{a} \hat{b}} + \frac{1}{2 \pi \alpha'} \nabla \widetilde{B}_{\hat{a} \hat{b}})^2}{16 m \widetilde{\rho}} \right] \Bigg\} \nonumber\\
  = & \, \int \mathcal{D} A_a \, \textrm{exp} \Bigg\{2 \pi i \int d\tau A_a \partial_\tau X^a + \frac{i}{\ell} \int d^3 x \left[- \frac{g}{4} (\widetilde{F} + \frac{1}{2 \pi \alpha'} \widetilde{B})^2 - \frac{(\nabla \widetilde{F}_{\hat{a} \hat{b}} + \frac{1}{2 \pi \alpha'} \nabla \widetilde{B}_{\hat{a} \hat{b}} )^2}{16 m \widetilde{\rho}} \right] \Bigg\}\, ,
\end{align}
where we can drop the last term in the IR regime, and
\be
  (\widetilde{F} + \frac{1}{2 \pi \alpha'} \widetilde{B})^2 \equiv \eta^{ab} \eta^{cd} \left(\widetilde{F}_{ac} + \frac{1}{2 \pi \alpha'} \widetilde{B}_{ac} \right) \left(\widetilde{F}_{bd} + \frac{1}{2 \pi \alpha'} \widetilde{B}_{bd} \right)\, .
\ee
Here again $\widetilde{F}$ and $\widetilde{B}$ are fluatuations of $F$ and $B$ respectively. The full theory should be the combination of the (2+1)D action above with the one from the (3+1)D duality, which has the following expression in the IR regime:
\begin{align}\label{eq:OpenStringAction}
  Z & = \int D B_{\mu\nu} \, D A_a \, \textrm{exp} \Bigg[ \frac{i \eta}{2} \sum_i\int_{\Sigma_i} d \sigma d\tau\, \epsilon^{\alpha\beta} \partial_{\alpha} X^\mu \partial_{\beta} X^\nu B_{\mu\nu} - \frac{i g}{2} \int d^4 x\, h_3^2 \nonumber\\
  {} & \qquad\qquad\qquad\qquad\quad + i \eta \sum_j  \int_{\partial \Sigma_j} d\tau\, A_a \partial_\tau X^a - \frac{i g}{4\ell} \int d^3 x\, (\widetilde{F}+\widetilde{B})^2 \Bigg],
\end{align}
where $\eta=2\pi \hbar$, $\alpha,\, 
\beta \in \{\tau,\, \sigma\}$, and $\Sigma_i$ is the worldsheet spanned by the $i$-th vortex line with boundaries $\partial \Sigma_j$. The summation over $\partial \Sigma_j$ includes all the endpoints $X^{a}=X^{a}(\tau)$ of vortex lines 
attached to dark solitons. From the 2-form field $B_{\mu\nu}$, one can define a 3-form field strength:
\be
  H_{\mu \nu \lambda}\equiv \partial_{\mu}B_{\nu\lambda}+\partial_{\nu}B_{\lambda\mu}+\partial_{\lambda}B_{\mu\nu} = H^{0}_{\mu\nu\lambda}+h_{\mu\nu\lambda}\, ,
\ee
where $H^{0}_{\mu\nu\lambda}$ is the background field with $H^{0}_{123}=\rho_0$, and the fluctuations are given by $h_3^2 =h_{\mu\nu\lambda} h^{\mu\nu\lambda} / 6$ with the 4D effective metric $\eta_{\mu\nu} = \textrm{diag} \{-c_s^2,\, 1,\, 1,\, 1\}$ and the 
speed of sound $c_s = \sqrt{g \rho_0 / m}$. $\widetilde{F}$ and $\widetilde{B}$ are the 
fluctuations of $F$ and $B$ on the soliton plane respectively.

The first line of Eq.~\eqref{eq:OpenStringAction} is just the effective string action without boundaries discussed in Refs.~\cite{Zee, Franz, Gubser}, while the second line of Eq.~\eqref{eq:OpenStringAction} is the contribution from boundaries. We would like to emphasize that the effective action \eqref{eq:OpenStringAction} is invariant under the following gauge transformations (see e.g. Ref.~\cite{Zwiebach}):
\begin{align}
  B_{\mu\nu} & \rightarrow B_{\mu\nu} + \partial_\mu \Lambda_\nu - \partial_\nu \Lambda_\mu\, , \nonumber\\
  A_a & \rightarrow A_a - \Lambda_a\, ,\label{eq:GaugeSymm}
\end{align}
where $\Lambda_\mu$ are transformation parameters. This kind of gauge symmetry is also called the $\Lambda$-symmetry, which has been studied in many works, e.g. Ref.~\cite{SheikhJabbari}.

\section{Solitons and $D$-branes}\label{sec:soliton}

In Section~\ref{sec:scalar}, we have discussed the relativistic version of the Gross-Pitaevskii theory. This theory has the same time-independent dark soliton solution as the non-relativistic one, which can be mapped into an effective string theory, as discussed in Section~\ref{sec:duality}.

Since we are interested in the time-independent solutions, the Lorentz-violating term in the non-relativistic Gross-Pitaevskii theory will be irrelevant. Moreover, we shall consider the large noncommutativity limit, in which the kinetic term can actually be neglected with respect to the potential. For these reasons the difference between the relativistic commutative scalar field theory and the non-relativistic Gross-Pitaevskii theory does not affect our discussions, as long as one focuses on the time-indepedent solutions. This simple observation enables us to use the more convenient relativistic theory to discuss some crucial issues of solitons.

In this section, we first discuss the noncommutative version of the Abelian Higgs model and justify the identification of their soliton solutions with the $D$-branes of the effective string theory introduced before. Our identification turns out to be valid in some limits, which we highlight by reviewing some crucial aspects of the noncommutative tachyon field theory appearing in the string theory literature \cite{Harvey,Banerjee:2004cw,Kim:2004xn}.

\subsection{Noncommutative Abelian Higgs Model}\label{NcAHM}

The commutative Abelian Higgs model has been discussed in Section~\ref{sec:scalar}, and it is given by the Lagrangian \eqref{eq:commAHM}. We are interested in the limit where the $U(1)$ gauge field is vanishing, and in this limit we obtain a scalar field theory given by Eq.~\eqref{eq:LcommScalar}. We consider several kinds of soliton solutions to this scalar field theory, in particular dark solitons \eqref{eq:darksoliton} and vortex lines \eqref{eq:VortexLine}.  In order to interpret these solitons from string theory's point of view, we will consider their counterparts in the noncommutative scalar field theory obtained from the noncommutative Abelian Higgs model in the limit of vanishing gauge field. This is the same strategy described in Ref.~\cite{AHM}.

As discussed in Ref.~\cite{SW}, the Seiberg-Witten map for the gauge field should preserve the gauge transformation relation, i.e.,
\be\label{eq:noncommConstr}
  \hat{A} (A) + \hat{\delta}_{\hat{\lambda}} \hat{A} (A) = \hat{A} (A + \delta_\lambda A)\, ,
\ee
where the hat ($\hat{\phantom{a}}$) denotes the noncommutative fields, and the gauge transformation for the ordinary Yang-Mills theory is
\be
  \delta_\lambda A_i = \partial_i \lambda + i [\lambda,\, A_i]\, ,
\ee
while for the noncommutative Yang-Mills theory:
\be
  \hat{\delta}_{\hat{\lambda}} \hat{A}_i = \partial_i \hat{\lambda} + i \hat{\lambda} \star \hat{A}_i - i \hat{A}_i \star \hat{\lambda}\, ,
\ee
with the star product $\star$ between two noncommutative fields $f$ and $g$ defined by 
\be
f(x)\star g(x)=e^{\frac{i}{2}\theta^{\mu\nu}\frac{\partial}{\partial \xi^\mu}\frac{\partial}{\partial \zeta^\nu}}\left.f(x+\xi)g(x+\zeta)\right\vert_{\xi=\zeta=0}\,,
\label{eq:Moyal}
\ee
and $\theta^{\mu\nu}$ is given by
\be
  [x^\mu,\, x^\nu] = i \theta^{\mu\nu}\, .
\ee
Solving the constraint \eqref{eq:noncommConstr}, one obtains
\begin{align}
  \hat{A}_i (A) & = A_i - \frac{1}{4} \theta^{kl} \{A_k, \, \partial_l A_i + F_{lk} \} + \mathcal{O} (\theta^2)\, ,\nonumber\\
  \hat{\lambda} (\lambda, A) & = \lambda + \frac{1}{4} \theta^{kl} \{\partial_k \lambda,\, A_l \} + \mathcal{O} (\theta^2)\, .
\label{SW-A} 
\end{align}
Similarly, for the scalar field we can also require that the gauge transformation relation should be preserved under the Seiberg-Witten map for both the commutative and the noncommutative fields, i.e.,
\be
  \hat{\phi} (\phi) + \hat{\delta}_{\hat{\lambda}} \hat{\phi} (\phi) = \hat{\phi} (\phi + \delta_\lambda \phi,\, A + \delta_\lambda A)\, .
\ee
Solving this equation for the Abelian gauge field at the order $\theta$, one obtains \cite{AHM}:
\be
  \hat{\phi} = \phi - \frac{1}{2} \theta^{kl} A_k \partial_l \phi + \mathcal{O} (\theta^2)\, .
\ee

Now, we can turn off the gauge field in the Abelian Higgs models to obtain the commutative and the noncommutative scalar field theory as follows:
\begin{align}
  \mathcal{L}_\textrm{c} & = -\frac{1}{2} \overline{(\partial_\mu \phi)} (\partial^\mu \phi) - \lambda (|\phi|^2 -v^2
   )^2\, ,\label{eq:commTachyon}\\
  \mathcal{L}_\textrm{nc} & = -\frac{1}{2} \overline{(\partial_\mu \phi)} \star (\partial^\mu \phi) - \lambda (\overline{\phi} \star \phi -v^2
   )^2\, ,\label{eq:noncommTachyon}
\end{align}
where the subscript ``c'' and ``nc'' stand for the commutative and the noncommutative theory respectively. The commutative theory is the same as the commutative scalar field theory \eqref{eq:LcommScalar}, which can be viewed as the relativistic version of the Gross-Pitaevskii theory \eqref{eq:LGP}, while the noncommutative one is just the noncommutative tachyon field theory that we are looking for. The Seiberg-Witten map for the scalar is trivial when turning off the gauge field:
\be
  \hat{\phi} = \phi\, .\
 \label{eq:TrivialMap}
\ee
However, the noncommutative scalar field still obeys the star product. The terms in \eqref{eq:noncommTachyon} containing one single star product can be replaced by the corresponding ones with an ordinary product, because the difference is only given by total derivatives, which vanish after the spacetime volume integration. Therefore, star products in the quartic interaction term are the only difference between the ordinary action \eqref{eq:commTachyon} and noncommutative field theory action \eqref{eq:noncommTachyon}. Our expectation is that the nonlocality introduced in this way may capture some stringy behavior, in particular when static noncommutative solitons are considered.

Noncommutative codimension-two soliton solutions have been found in the large noncommutativity limit in Ref.~\cite{GMS}. In the large noncommutativity limit, the kinetic term can be neglected and the solitons need to satisfy
\be
  \frac{dV_\star}{dT} = 0\, ,
\label{eq:SolEq}  
\ee
where $V_\star(T)=\lambda (\overline{T} \star T +v T+ v \overline{T} )^2$ is the potential expressed in terms of the shifited scalar field $T=\phi-v$. We restrict our discussions to real solutions. To find a such solution, one needs to first find a field $\phi_0$ satisfying
\be\label{eq:starCond}
  \phi_0 \star \phi_0 = \phi_0\, ,
\ee
and then any function $F$ with the form $F(x) = \sum_{n=1}^\infty a_n x^n$ has the property
\be
  F (\lambda \phi_0) = F(\lambda)\, \phi_0\, .
\ee
Hence, a potential of this form $V(T)$ obeys
\be
  \frac{dV}{dT} \bigg|_{T=\lambda \phi_0} = \left(\frac{dV}{dT} \bigg|_{T=\lambda} \right) \, \phi_0\, .
\ee
The simplest function satisfying the condition \eqref{eq:starCond} is given by \cite{GMS}:
\be
  \phi_0 (r) = 2 \, e^{-r^2 / \theta}\, ,\quad r^2 = x_1^2 + x_2^2\, ,
\ee
where $\theta \equiv \theta^{12}$ in the large noncommutativity limit.
The solution to equation \eqref{eq:SolEq} takes the form
\be
T=-v \phi_0 (r)\,.
\label{eq:Co2Soliton}
\ee
Note that this solution correctly interpolates between the value $T=0$ where $V(T)$ has a local minimum and $T=-v$ corresponding to an unstable local maximum.
This construction can be generalized to arbitrary even-codimension solitons by replacing $r^2 = x_1^2 + x_2^2$ with $r^2 = x_1^2 + x_2^2 + \cdots + x_{2q}^2$. One can easily infer that a codimension-two soliton is just the noncommutative counterpart of the modulus of the vortex line solution \eqref{eq:VortexLine}.

Now we want to look into the possibility of identifying a such noncommutative solution with a $D(p-2)$-brane coming from the decay of an unstable $Dp$-brane, and more generally we would like to relate a codimension-$2q$ soliton with a $D(p-2q)$-brane. A crucial test is the value of the soliton energy, which can be easily calculated, since in the large noncommutativity limit one can neglect the derivative term in the tachyon effective action \eqref{eq:noncommTachyon}, namely,
\be
  S = - \int d^{p+1} x\, V_\star(T)\, .
\ee
Inserting the soliton solution \eqref{eq:Co2Soliton} into the equation above and integrating over $x_1$ and $x_2$, we get
\be
  S = -V(-v) \int d^{p-1} x\, \int d^2 x\, \phi_0 (r) = -2 \pi \theta \, V(-v)\int d^{p-1} x\, ,
\ee
from which we can read off the tension $\mathcal{T}_{p-2}=2 \pi \theta \, V(-v)=2 \pi \theta \, \lambda v^4$. On the other hand,  the constant vacuum energy of the action \eqref{eq:noncommTachyon} gives the value of the tension of the unstable $Dp$-brane
\be
\mathcal{T}_p=V(-v)=\lambda v^4\, .
\ee 
Therefore, we obtain the descent relation
\be
\mathcal{T}_p = (2\pi \theta)\, \mathcal{T}_{p-2}\, ,
\ee
which is to be compared with the expected relation $\mathcal{T}_p = (2\pi )^2\alpha'\, \mathcal{T}_{p-2}$, where $\alpha'=-1/{m_T^2}=1/{m_H^2}=1/(2\sqrt{\lambda}v)$.

One can also construct codimension-one solitons. For a noncommutative field $\hat\phi(x)$ depending only on one coordinate $x$, such that $\partial_t \hat\phi(x)=\partial_{y^a}\hat\phi(x)=0$ for $y^a\neq x$, one can easily show that
\be
\hat{\phi}(x)\star\hat{\phi}(x)=\hat{\phi}^2 \,.
\ee
As this product is the only place where noncommutativity appears in the action \eqref{eq:noncommTachyon}, the same procedure that was used to get the commutative soliton \eqref{eq:darksoliton} can be used to prove that it is also a solution in the noncommutative case. However, interpreting a codimension-one soliton as a $D(p-1)$-brane from the decay of an unstable $Dp$-brane, the following descent relation for tensions has been derived in Refs.~\cite{Banerjee:2004cw,Kim:2004xn}:
\be
\mathcal{T}_p = \frac{8\sqrt{2}}{3}\frac1{m_H}\, \mathcal{T}_{p-1}\, ,
\ee
which differs from the expected one shown above. We will make more discussions on this puzzle in the next subsection.

\subsection{Noncommutative Tachyon Field Theory}\label{noncommTFT}

In the previous subsection, we have seen a noncommutative scalar field theory obtained by a generic noncommutative mapping of the relativistic Gross-Pitaevskii, whose soliton solutions correspond to lower-dimensional branes in string theory. In order to set up the correspondence at quantitative level, we need to investigate the relation between the noncommutative scalar theory and the noncommutative tachyon theory, that is obtained from string theory. The latter is known to have solitons reproducing the correct tensions for $D$-branes. We will see that the crucial ingredient to get them is to consider the specific kind of Seiberg-Witten map prescribed by string theory.

First of all, we discuss the connection between the effective open string theory discussed in the previous section and the corresponding tachyon field theory. The effective open string theory \eqref{eq:OpenStringAction} contains the standard bulk action of string theory in a background $B$-field, namely
\begin{equation}
S_{0}=\frac{1}{4\pi\alpha'}\int_{\Sigma}\left(g_{\mu\nu}\partial_{a}X^{\mu}\partial^{a}X^{\nu}-2\pi i\alpha'B_{\mu\nu}\epsilon^{ab}\partial_{a}X^{\mu}\partial_{b}X^{\nu}\right)\, ,
\label{eq:SA}
\end{equation}
where $\mu,\nu=0,1,\ldots,D$, with $D$ denoting the dimension of the space-filling $D_{(D-1)}$-brane, and $B_{\mu\nu}$ is an antisymmetric matrix of rank $r=4$, so that we can assume $B_{\mu\nu}\neq 0$ only for $\mu,\nu=0,\ldots,3$. On top of this we take $B_{0\nu}=0$ for any $\nu$. We will also assume $g_{\mu\nu}=0$ for $\mu=0,\ldots,3$ and $\nu \neq 0,\ldots, 3$. Therefore, we have to take $D\geq 4$ and in particular we can choose $D=26$, corresponding to critical strings. Furthermore, we assume constant background fields, i.e. $g_{\mu\nu}\approx constant$ and $B_{\mu\nu}\approx constant$, such that $H_{\mu\nu\rho}\approx 0$. In the action \eqref{eq:SA}, we have included the standard kinetic term, which was not present in the action \eqref{eq:OpenStringAction}, because it is suppressed in the large noncommutativity limit:
\be
\alpha' B_{\mu\nu} \to \infty\, ,\quad g_{\mu\nu}\,\, \textrm{fixed}\, ,
\ee
with $\mu, \nu = 0, \ldots, 3$, or equivalently $g_{\mu\nu} / \alpha' \to 0$ while keeping $B_{\mu\nu}$ fixed. The directions $\mu=4,\ldots,25$ in the action \eqref{eq:SA} can be treated rather trivially, contributing just an overall factor to the corresponding partition function. As we saw in the previous section, open strings are characterized by a worldsheet with the topology of a disk, so we can in general introduce a term
\be
S'
=\intop_{0}^{2\pi}\frac{d\sigma}{2\pi}\mathcal{V}\, ,
\ee
where $\sigma\in \left[0,2\pi \right)$ is a parameter on the boundary $\partial \Sigma$, and $\mathcal{V}$ is a general boundary perturbation. In particular, in the previous section we considered
\begin{equation}
\mathcal{V}=-i A_\mu (X) \partial_\sigma X^\mu  \,,  
\label{eq:Abp}
\end{equation}
 where $A_\mu (X^0,\ldots,X^3) $  is a $U(1)$ gauge field. Using the Stokes theorem, for a slowly varying $A_\mu$ or a constant $F_{\mu\nu}$ this boundary  term can be written as 
\begin{equation}
S'=-\frac{i}{2}\intop_{0}^{2\pi}F_{\mu\nu}X^{\mu}\partial_\sigma X^\nu\,.
\end{equation}
 Analogously, for a constant $B$-field we can rewrite 
 \begin{equation}
 -\frac{i}{2}\int_{\Sigma}B_{\mu\nu}\epsilon^{ab}\partial_{a}X^{\mu}\partial_{b}X^{\nu}=
 -\frac{i}{2}\intop_{0}^{2\pi}B_{\mu\nu}X^{\mu}\partial_\sigma X^\nu\,.
 \end{equation}
In this section, we are interested in the scalar effective action that can be obtained by considering the slowly varying boundary deformation $T(X^0,X^1,X^2,X^3)$ along the directions, in which $B$ is non-null. In the infrared (IR) approximation under consideration, where all non-convariant terms are neglected, this tachyon theory will be relativistic. In principle, it should include terms with derivatives of arbitrary order, which makes a direct study of its solitons very daunting. On the other hand, in the large noncommutativity limit the theory is dramatically simplified and its solitons can be related to the ones studied for the Gross-Pitaevskii scalar theory in the previous subsection.

The noncommutative tachyon field theory has its origin in string theory in the presence of a constant $B$-field. In Ref.~\cite{SW} it has been shown that, when the boundary perturbation \eqref{eq:Abp} is considered with a slowly varying $A_\mu$, one can obtain an effective action of the form 
\begin{equation}\label{ncbi}
{\hat S}_{{\rm DBI}}= -\frac{1}{G_{{\rm
s}}(2\pi)^p(\alpha')^{\frac{p+1}{2}}}\int d^{p+1}x \sqrt{-{\rm det}
(G_{\mu\nu}+2\pi\alpha'(\hat{F}_{\mu\nu}+\Phi_{\mu\nu}))}\,,
\end{equation}
where the open string metric $G_{\mu\nu}$, the open string coupling $G_s$  and the two-form $\Phi$  are determined by the formulae
\begin{eqnarray}
&&\frac{1}{G + 2\pi\alpha'\Phi} =-\frac{\theta}{2\pi\alpha'} + \frac{1}{g + 2\pi\alpha'B}\, , \label{ocl1} \\
&& G_s = g_{{\rm s}} \sqrt{\frac{\det (G + 2\pi\alpha'\Phi)}{\det (g +
2\pi\alpha'B)}}\, . \label{ocl2}
\end{eqnarray}
In the first equation \eqref{ocl1}, $G$ and $\Phi$ are determined in terms of the closed string metric $g$, the closed string coupling $g_s$, $B$ and an arbitrary noncommutativity parameter $\theta$, because they are symmetric and antisymmetric respectively. The gauge field $\hat A$ is related to the commutative $A$ by Eq.~\eqref{SW-A}. The second equation \eqref{ocl2} is motivated by demanding that the effective action \eqref{ncbi} with $\hat{F}=0$ is the same as the usual commutative Dirac-Born-Infeld (DBI) action with $F=0$. Actually, it has been shown that, using the transformations of the fields $\hat{A}(A)$ given by Eq.~\eqref{SW-A}, the two Lagrangians are related as
\begin{equation}\label{rel}
 {\cal L}_{{\rm DBI}}= \hat {\cal L}_{{\rm DBI}} + (\mbox{total}~~\mbox{derivative}) + {\cal O}(\partial F)\, .
\end{equation}
For $\theta=0$, Eq.~\eqref{rel} is obvious. For
\be
  \theta^{\mu\nu} = - (2 \pi \alpha')^2 \, \left(\frac{1}{g + 2 \pi \alpha' B} B \frac{1}{g - 2 \pi \alpha' B} \right)^{\mu\nu}\, .
\ee
the interpolating field $\Phi$ vanishes.

In Ref.~\cite{Banerjee:2004cw, Kim:2004xn} the same argument was used for the DBI-like effective action of the tachyon field for an unstable $Dp$-brane:
\be
\label{fa}
S=-\frac{1}{g_{{\rm s}}(2\pi)^p(\alpha')^{\frac{p+1}{2}}} \int
d^{p+1}x V(T) \sqrt{-{\rm det} (g_{\mu\nu}+2\pi\alpha'(B_{\mu\nu}
+\partial_\mu T \partial_\nu T))}\, ,
\ee
where $V(T)$ measures the variable tension of the unstable $D$-brane, and is a runaway potential, monotonically connecting its maximum coinciding with the tension of the $Dp$-brane and its minimum representing the vanishing unstable $Dp$-brane. In particular, one can choose conventions in which $V(T=0)=1$ and $V(T=\pm\infty)=0$. We can therefore read off
\be
\mathcal{T}_p=\frac{1}{g_{{\rm s}}(2\pi)^p(\alpha')^{\frac{p+1}{2}}}\,.
\ee 
If we ignore terms with higher-order derivatives of the tachyon, the corresponding noncommutative tachyon action can be written as  
\begin{equation}\label{ncfa}
\hat{S}= -\hat{\cal T}_p \int d^{p+1}x\;
{V_\star}(\hat{T})  \sqrt{ -{\rm  det}\left(G_{\mu\nu}+2\pi\alpha'\partial_\mu \hat{T} \partial_\nu \hat{T}\right) } \, ,
\end{equation}
where $\hat{T}$ is related to $T$ in the trivial way \eqref{eq:TrivialMap}, $V_\star(T)$ is obtatined from $V(T)$ by replacing ordinary products with $\star$-products, and 
\be
\hat{\mathcal{T}}_p=\frac{1}{G_{{\rm s}}(2\pi)^p(\alpha')^{\frac{p+1}{2}}}=\frac{\mathcal{T}_p}{\sqrt{\det (1 +
2\pi\alpha' g^{-1} B)}}\,.
\ee
What is remarkable about the action \eqref{fa}, which is written in commutative formalism, is that the Seiberg-Witten map replaces it with the completely equivalent noncommutative expression \eqref{ncfa}. This is different from the case discussed in Subsection~\ref{NcAHM}, where the noncommutative theory was different from its commutative counterpart. Nevertheless, it is not hard to see that, for slowly varying tachyons such that $\partial_\mu T$ is small, the actions \eqref{fa} and \eqref{ncfa} reduce to the more common tachyon actions of the form \cite{Tseytlin, Okuyama:2000ch}:
\be
  S = \frac{C}{g_s} \int d^{p+1} x\, \mathcal{L}_{BI}(B) \left[\frac{1}{2} f(T) g^{ij} \partial_i T\, \partial_j T - V(T) + \cdots \right]\, ,
\ee
and
\be\label{eq:tachyonAction}
 \hat{S} = \frac{C}{G_s} \int d^{p+1} x\, \sqrt{G} \left[\frac{1}{2} f_\star(T) G^{ij} \partial_i T\, \partial_j T - V_\star(T) + \cdots \right]\, ,
\ee
where $C=g_s\, \mathcal{T}_{D_p}$, and $\mathcal{L}_{BI}(B)=\sqrt{\det(g+2\pi\alpha'B)}$ is the usual DBI action for the vanishing gauge field. The factor $f_\star(T)$, like $V_\star(T)$, is simply obtatined from $f(T)$ by replacing ordinary products with $\star$-products.

Several computations of the tachyon potential are available in the literature using some techniques from string field theory. To simplify the calculations, one can also consider the large noncommutativity limit. We will present some details in Appendix~\ref{app:TachyonPot}, and the result for bosonic string theory is \cite{WittenTachyon-1, WittenTachyon-2, Shatashvili-3, KutasovMarino,Cornalba:2000ad,Okuyama:2000ch}:
\be\label{eq:TachyonPot-1}
  V(T) = (T+1)\, e^{-T}\, .
\ee
It is explained in Ref.~\cite{Shatashvili-3}, that although suppressed the kinetic term in the action does not take a canonical form in terms of the tachyon field $T$, and a change of variable $\textrm{exp} (-T) = \phi^2$ will lead to the canonical kinetic term, and correspondingly the tachyon potential can be expressed in the new variable as
\be\label{eq:TachyonPot-2}
  V(\phi) = - \phi^2 \, \textrm{log} \frac{\phi^2}{e}
\ee
in Euclidean signature. One can expand it around the vacuum $\phi^2 = 1$. When only real fields are considered, the leading order expression of the potential in Minkowski signature reproduces the shape of the potential in the noncommutative tachyon field theory \eqref{eq:noncommTachyon} that has been discussed in Subsection~\ref{NcAHM}.

Now let us return to the effective string action \eqref{eq:OpenStringAction} that is obtained from the non-relativistic Gross-Pitaevskii theory using the boson/vortex duality. Since there is a gauge symmetry \eqref{eq:GaugeSymm}, we can choose a gauge $A_a = 0$. Moreover, we focus on the region around the dark solitons, where the $B$-field can be approximately viewed as constant, which implies that the field strength terms $\sim h_3^2$ and $\sim (\widetilde{F} + \widetilde{B})^2$ in Eq.~\eqref{eq:OpenStringAction} can be neglected near dark soliton planes. Also, as discussed in Section~\ref{sec:duality}, a term violating the Lorentz symmetry can be dropped in the IR regime. Hence, under these approximations the effective action \eqref{eq:OpenStringAction} is exactly the same as the standard string action in the large $B$-field limit, and the analysis above holds in our case. It means that near dark soliton planes the effective string theory \eqref{eq:OpenStringAction} can provide an effective noncommutative tachyon field theory of the form \eqref{eq:tachyonAction} with the tachyon potential \eqref{eq:TachyonPot-1} or \eqref{eq:TachyonPot-2}, which at the leading order coincides with the theory \eqref{eq:noncommTachyon} obtained from the noncommutative Abelian Higgs model.

We can consider the same kind of codimension-two solitons discussed in the previous subsection, namely $T_*\, \phi_0 (r)$, where $T_*$ is the value at which $V(T)$ has a local maximum. The tachyon potential $V_\star$  for such a solution has the value $V(T_*)\, \phi_0 (r)$, where $V(T_*) = 1$ in our conventions. Therefore,
\be
  S = -\frac{C\, V(T_*)}{G_s} \int d^{p-2} x\, \int d^2 x\, \sqrt{G}\, \phi_0 (r) = -\frac{2 \pi \theta C\, V(T_*)}{G_s} \int d^{p-2} x\, \sqrt{G}\, .
\ee
From Eq.~\eqref{ocl2} we obtain for the large $B$-field the relation between the open string coupling $G_s$ and the closed string coupling $g_s$:
\be
  G_s = \frac{g_s \sqrt{G}}{2 \pi \alpha' B \sqrt{g}}\, ,
\ee
where $B=B_{12}$. Taking into account $\theta = 1/B$, we obtain
\be
  S = - (2\pi)^2 \, \alpha' \frac{C}{g_s} \int d^{p-2} x\, \sqrt{g}\, .
\ee
Hence, the soliton tension is
\be
 \mathcal{T}_{\textrm{sol}} = (2\pi)^2 \alpha' \frac{C}{g_s} = (2\pi)^2 \alpha' \mathcal{T}_{p} = \mathcal{T}_{p-2}\, ,
\ee
where $C=\mathcal{T}_p g_s$. This result is consistent with the one for the bosonic $D(p-q)$-branes:
\be\label{eq:D-braneTension}
  \mathcal{T}_{p-q} = (2\pi \sqrt{\alpha'})^{q}\, \mathcal{T}_p\, .
\ee

We can also consider codimension-one soliton solutions, whose energy is expressed as a one-dimensional integral in Eq.~\eqref{eq:DarkSolEnergy}. As discussed in Ref.~\cite{BEC}, to prevent the long-wavelength instabilities we have to restrict the sizes of the transverse dimensions to be $\sim 2 \pi \sqrt{\alpha'}$, where $\ell_s = \sqrt{\alpha'}$ is the characteristic length of the effective strings, i.e. the length of the vortex lines. Hence, we expect that the relation \eqref{eq:D-braneTension} also holds for $q=1$, i.e. the codimension-one soliton solutions. On the other hand, it has been computed in Refs.~\cite{Banerjee:2004cw,Kim:2004xn} using the tachyon field theory, that the relation for the tension of codimension-one soliton solutions coincides with the one for $D(p-1)$-branes:
\begin{displaymath}
 \mathcal{T}_{p-1} = (2\pi \sqrt{\alpha'})\, \mathcal{T}_p\, ,
\end{displaymath}
which also supports the identification of dark solitons in Gross-Pitaevskii theory and $D(p-1)$-branes in the effective string theory.

\subsection{$D$-brane Interaction}\label{check}

As another check of identifying dark solitons with $D$-branes, in this subsection we compute the interaction between two parallel $D$-branes in the effective string theory. A well-known computation in the ordinary string theory has been done by Polchinski in Ref.~\cite{Polchinski} (see e.g. Ref.~\cite{D-braneBook} for a summary).

\begin{figure}[!htb]
      \begin{center}
        \includegraphics[width=0.5\textwidth]{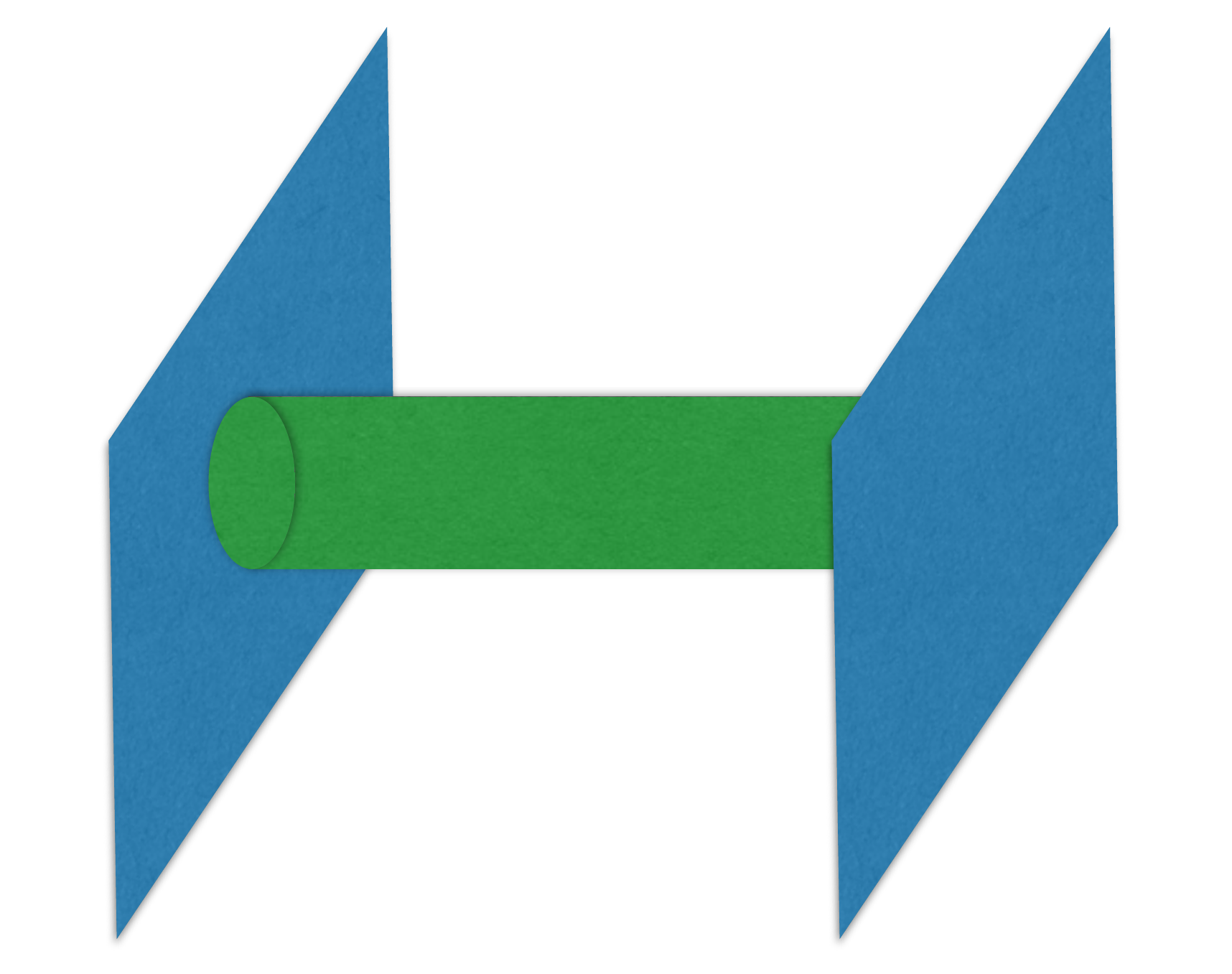}
      \caption{The exchange of a closed string between two parallel $D$-branes}
      \label{fig:D-braneInt}
      \end{center}
    \end{figure}

The calculation is essentially to evaluate the amplitude of exchanging a closed string between two parallel $D$-branes (see Fig.~\ref{fig:D-braneInt}), or equivalently to evaluate a 1-loop amplitude of open strings. For the picture of closed strings in the NS-NS sector, only the graviton and the dilaton were taken into account, because the antisymmetric $B$-field contributes at higher order. If one considers the type-II superstring, the contribution from the R-R sector will exactly cancel the one from the NS-NS sector, as discussed in Ref.~\cite{Polchinski}. However, fixing the gauge $A_a=0$ and neglecting the fluctuations, the effective string theory \eqref{eq:OpenStringAction} can be thought of as the large $B$-field limit of the ordinary string theory. Hence, in this case the contributions from the graviton and the dilaton should be neglected, and only the $B$-field contributes to the potential between two parallel $D$-branes.

To compute the amplitude of exchanging the $B$-field between two parallel $D$-branes, we need the propagator of the $B$-field in the bulk and the coupling between the $B$-field and the $D$-brane. Following Ref.~\cite{D-braneBook}, we can read off the bulk propagator of the $B$-field from the effective action in the Einstein frame:
\begin{align}
  S^E & = \frac{1}{2 \kappa^2} \int d^D x\, (- \widetilde{G})^{1/2} \Bigg[\widetilde{R} - \frac{4}{D-2} \nabla_\mu \widetilde{\Phi} \nabla^\mu \widetilde{\Phi} - \frac{1}{12} e^{- 8 \widetilde{\Phi} / (D-2)} H_{\mu\nu\lambda} H^{\mu\nu\lambda} \nonumber\\
  {} & \qquad\qquad\qquad\qquad\qquad - \frac{2 (D-36)}{3 \alpha'} e^{4 \widetilde{\Phi} / (D-2)} + \mathcal{O} (\alpha') \Bigg]\, ,
\end{align}
where
\begin{align}
  \widetilde{\Phi} & = \Phi - \Phi_0\, ,\nonumber\\
  \widetilde{G}_{\mu\nu} & = e^{\frac{4 (\Phi_0 - \Phi)}{D-2}} G_{\mu\nu}\, ,
\end{align}
and $\widetilde{R}$ is the corresponding Ricci scalar after the transformation. The terms relevant to the bulk propagator of the $B$-field are
\be
  S^E \supset -\frac{1}{24 \kappa^2} \int d^D x\, B_{\mu\nu} \square B_{\mu\nu}\, ,
\ee
and the bulk propagator of the $B$-field in momentum space is \cite{Townsend}:
\be
  \langle B_{\mu\nu} \, B_{\rho\sigma} \rangle = - \frac{6 i \kappa^2}{k^2} (\delta_{\mu\rho} \, \delta_{\nu\sigma} - \delta_{\nu\rho} \delta_{\mu\sigma})\, .
\ee
To obtain the coupling between the $B$-field and the $D$-brane, we expand the DBI-action in the Einstein frame:
\be
  S_p^E = - \tau_p \int d^{p+1} \xi\, e^{- \widetilde{\Phi}} \sqrt{\textrm{det} \left(e^{\frac{4 \widetilde{\Phi}}{D-2}} \, \widetilde{G}_{ab} + B_{ab} + 2 \phi \alpha' F_{ab} \right)}\, ,
\ee
where the indices $a$, $b$ run over the ($p+1$)-dimensions on the $D$-brane. The terms relevant to the coupling between the $B$-field and the $D$-brane are
\be
  S_p^E \supset - \frac{\tau_p}{4} \int d^{p+1} \xi\, B_{ab} \, B^{ab}\, .
\ee
From this coupling we see that the leading order contribution is already at 1-loop order, thus from field theory point of view we need to evaluate the following 1-loop graph:
    \begin{figure}[!htb]
      \begin{center}
        \includegraphics[width=0.5\textwidth]{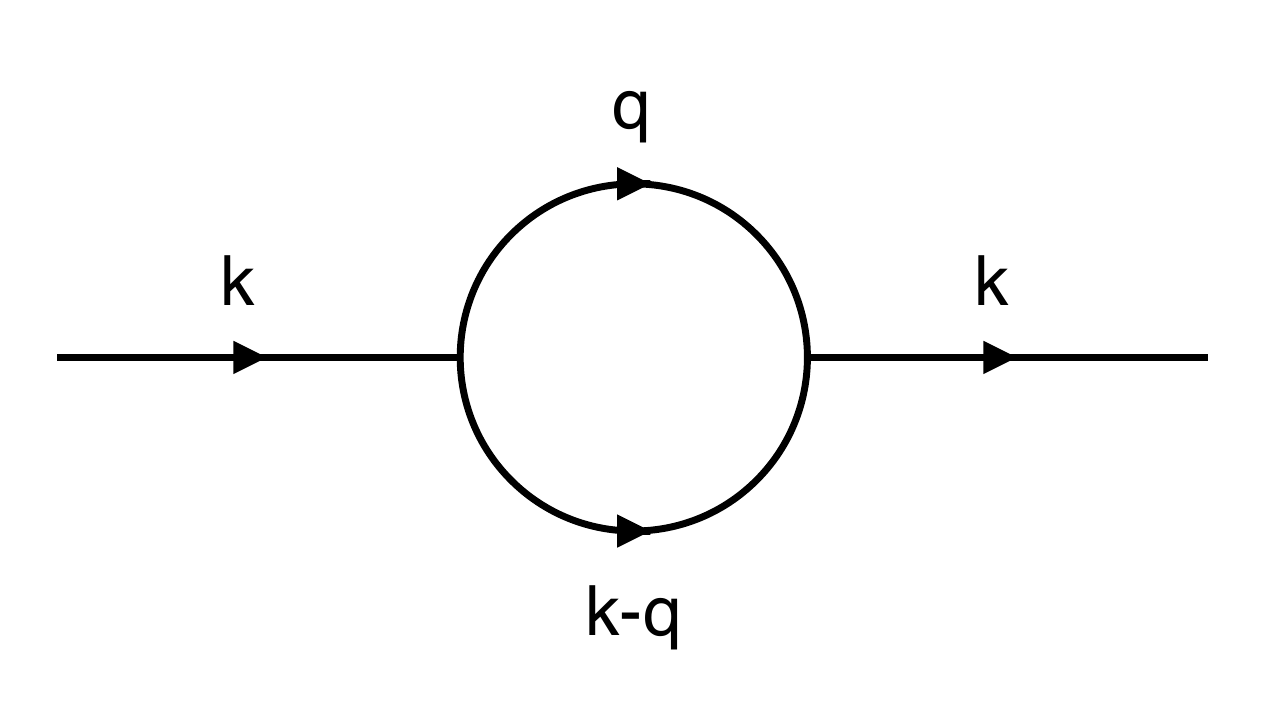}
      \caption{The 1-loop Feynman diagram of the $B$-field coupled to $D$-branes}
      \label{fig:1-loop}
      \end{center}
    \end{figure}\\
The amplitude is
\begin{align}
  i \mathcal{M} (k) & = \frac{1}{2} \left(-\frac{i \tau_p}{2} \right)^2 \int \frac{d^{D-p-1} q}{(2 \pi)^{D-p-1}}\, \left(- \frac{6 i \kappa^2}{q^2} \right) \, \left(- \frac{6 i \kappa^2}{(k-q)^2} \right) \nonumber\\
  {} & \qquad\qquad\qquad \cdot \frac{\partial X^\mu}{\partial \xi^a}\, \frac{\partial X^\nu}{\partial \xi^b}\, \frac{\partial X^\rho}{\partial \xi^c}\, \frac{\partial X^\sigma}{\partial \xi^d}\, (\delta_{\mu\rho} \, \delta_{\nu\sigma} - \delta_{\nu\rho} \delta_{\mu\sigma}) \nonumber\\
  {} & \qquad\qquad\qquad \cdot \frac{\partial X^{\bar{\mu}}}{\partial \xi^a}\, \frac{\partial X^{\bar{\nu}}}{\partial \xi^b}\, \frac{\partial X^{\bar{\rho}}}{\partial \xi^c}\, \frac{\partial X^{\bar{\sigma}}}{\partial \xi^d}\, (\delta_{\bar{\mu} \bar{\rho}} \, \delta_{\bar{\nu} \bar{\sigma}} - \delta_{\bar{\nu} \bar{\rho}} \delta_{\bar{\mu} \bar{\sigma}}) \nonumber\\
  {} & = \frac{9}{2} \tau_p^2 \kappa^4 \left[(p+1)^2 - 2 (p+1) \right] \, \int \frac{d^{D-p-1} q}{(2 \pi)^{D-p-1}}\, \frac{1}{q^2 (k-q)^2} \nonumber\\
  {} & = \frac{9}{2} \tau_p^2 \kappa^4 (p^2 - 1) \frac{i^{D-p-4}}{(4 \pi)^{\frac{D-p-1}{2}}} \frac{\Gamma \left(\frac{5-D+p}{2} \right)\, \Gamma \left(\frac{D-p-3}{2} \right)^2}{\Gamma (D-p-3)} \, \frac{1}{k^{5-D+p}}\, .
\end{align}
To obtain the potential in the spacetime, we should apply the Born approximation and Fourier transform the amplitude $- \mathcal{M} (k)$.

In order to compare with the interactons between dark solitons, we consider the case $D=4$, $p=0$. As we discussed before, to make the dark soliton relatively stable and compatible with the Derrick's theorem, one has to restrict the size of transverse dimensions on the $D$-brane, i.e. to confine the system in a cylindrical geometry. Hence, in this computation the $D$-branes in real BEC systems can be effectively thought of as D0-branes. After the Fourier transform of the amplitude, we obtain the potential between two parallel $D$-branes ($D=4$, $p=0$):
\begin{align}
  V(x) & = i \frac{9}{16} \tau_p^2 \kappa^4 \int \frac{d^3 k}{(2 \pi)^3} \frac{e^{i k \cdot x}}{k} \nonumber\\
  {} & = \frac{9\, \tau_p^2 \kappa^4}{32 \pi}\, \frac{\delta(x)}{x}\, .\label{eq:D-branePot}
\end{align}
We see that, strictly speaking the contribution of the $B$-field to the interaction between two D0-branes is given by a Dirac $\delta$-function, i.e. a contact interaction. However, in reality the size of transverse dimensions on the $D$-brane is not zero, although negligible compared to the distance between two $D$-branes. Hence, the Dirac $\delta$-function in Eq.~\eqref{eq:D-branePot} can be understood as
\be
  \lim_{\ell \to 0} \frac{e^{-x^2 / \ell^2}}{\sqrt{\pi} \ell}\, ,
\ee
where $\ell$ is proportional to the size of transverse dimensions on the $D$-brane. Therefore, we expect the potential between two parallel dark solitons in BEC systems is a short-ranged repulsion and exponentially decaying. As far as we know, there is no analytical expression of the potential between two parallel dark solitons available in the literature, and the numerical results \cite{D-braneInt} are consistent with our results from string theory computation at qualitative level. More interestingly, some recent studies \cite{DarkSolitonAttraction} in optical systems confirmed experimentally that, dark solitons can have attractions only when some nonlocal response is turned on, which is also consistent with our expectation from string theory, i.e., in the presence of a string tension term in the action, the exchange of the graviton and the dilaton will induce an attrative interaction between two parallel $D$-branes.

\section{Discussions}\label{sec:discussion}

In this paper we have discussed the duality map between the Gross-Pitaevskii theory and a (3+1)D effective string theory. We generalize the previous works \cite{Zee, Gubser} to the spacetime with boundaries (see also Ref.~\cite{BEC}). As a consequence, we identify dark soliton solutions in the Gross-Pitaevskii theory and $D$-branes in the effective string theory under certain approximations, and various checks have been made to test this identification. With this new perspective, one has an opportunity to test many results and predictions of string theory in real experiments and on the other hand bring in new ideas to the study of quantum fluids and cold atom systems.

We would like to explore more aspects of this duality and its relation to a real cold atom system at quantitative level. For instance, Ref.~\cite{BEC} has started discussing the stability of the configuration of open vortex lines attached to the dark solitons, and we believe that a more detailed analysis of this dual picture can help us study the time evolution of $D$-brane decay. More interestingly, by introducing some fermionic fields an emergent supersymmetry can be realized in the cold atom systems. We hope that this can help stabilize dark solitons, in the same way of stabilizing $D$-branes from the superstring theory, and eventually to help simulate superstring theory in real experiments.

From more theoretical point of view, the boson/vortex duality discussed in this paper is also of great interest. As mentioned in Section~\ref{sec:duality}, the duality can be generalized to other dimensions. Since the (1+1)D Gross-Pitaevskii equation, also called the nonlinear Schr\"odinger equation, is an integrable model, we expect the integrability should be maintained in the dual theory \cite{Integ-2}. Also, the (1+1)D nonlinear Schr\"odinger equation is dual to a 2D topological Yang-Mills-Higgs model at quantum level \cite{GS}. By constructing the gravity dual of the 2D topological Yang-Mills-Higgs model, we expect that the $D$-branes in the supergravity theory correspond to the soliton solutions of the (1+1)D nonlinear Schr\"odinger equation. These results will be presented elsewhere \cite{Integ-1}.

More mathematically, the identification of solitons and $D$-branes discussed in this paper can also be understood from the viewpoint of K-theory. As discussed in Ref.~\cite{D-braneBook}, in the annihilation of a $D_p$-brane and an anti-$D_p$-brane, if the tachyon field is given by a topologically stable kink depending only on one of the dimensions inside the brane, then a $D_{(p-1)}$-brane will be left over after the annihilation. In our case, the (anti-)$D_p$-brane can be viewed as the space-filling (anti-)$D_3$-brane, while in the end we should see a $D_2$-brane left, which can be identified as the dark soliton in the Gross-Pitaevskii theory. More details of this perspective and its applications to topological phases will be explored in the future work.

In stead of the boson/vortex duality, some recent works \cite{KarchTong, MuruganNastase, SeibergWitten} have discussed a closely related particle/vortex duality web, especially the dual of the fermionic field theory in (2+1)D. To apply these ideas to the (3+1)D Abelian Higgs model and understand the corresponding web of dualities will help us understand the phase transition, the vacuum structure and the renormalization group flow of the theory, which we would like to pursue soon.

\section*{Acknowledgements}

We would like to thank Loriano Bonora, Ilmar Gahramanov, Steven Gubser, Song He, Antonio Mu\~noz Mateo, Nikita Nekrasov, Vasily Pestun, Vatche Sahakian, Ashoke Sen and Xiaoquan Yu for many useful discussions. We also would like to thank M. M. Sheikh-Jabbari and J. Murugan for communications.

\appendix

\section{Solutions to the Gross-Pitaevskii Equation}\label{app:sol}

In this appendix we briefly summarize the solutions with nontrivial topology to the Gross-Pitaevskii theory \eqref{eq:LGP}.

\subsection*{Dark Soliton}

The classical soliton solutions can be found by solving the Gross-Pitaevskii equation directly. A more field-theoretic way of finding the soliton solutions is to use the standard BPS approach, which we will briefly review now.

Based on the famous Derrick's theorem, a stable soliton solution for the pure scalar theory exists only for dimensions $D \leq 2$. Hence, we restrict our discussions to the (1+1)D solutions in the following, i.e., we assume that the soliton solutions are independent of two other spatial dimensions in (3+1)D. The BPS procedure for the (1+1)D scalar field theory can be summarized as follows.

A general scalar field theory is given by
\be
  \mathcal{L} = -\frac{1}{2} (\partial_x \phi)^2 - V(\phi)\, ,
  \label{RealScalar}
\ee
which leads to the field equation
\be\label{eq:FieldEquation}
  \partial_x^2 \phi - V'(\phi) = 0\, .
\ee
If the potential $V(\phi)$ can be expressed as
\be
  V = (W')^2\, ,
\ee
the energy of the system is given by
\be
  E = \int_{-\infty}^\infty dx\, \left[\frac{1}{2} (\partial_x \phi)^2 + (W')^2 \right]\, ,
\ee
where $W$ is a functional of the field $\phi$, and
\be
  W' \equiv \frac{\partial W}{\partial \phi}\, .
\ee
Consequently,
\begin{align}
  E & = \int_{-\infty}^\infty dx\, \left[\left(\frac{1}{\sqrt{2}} \partial_x \phi - W' \right)^2 + \sqrt{2}\, W' \partial_x \phi \right] \nonumber\\
  {} & = \int_{-\infty}^\infty dx\, \left[\left(\frac{1}{\sqrt{2}} \partial_x \phi - W' \right)^2 + \sqrt{2}\, \frac{\partial W}{\partial x} \right] \nonumber\\
  {} & = \int_{-\infty}^\infty dx\, \left[\left(\frac{1}{\sqrt{2}} \partial_x \phi - W' \right)^2 \right] + \sqrt{2} \left[ W(+\infty) - W(-\infty) \right]\, .\label{eq:BPSprocess}
\end{align}
If $W(+\infty)$ and $W(-\infty)$ correspond to different vacua, the configuration provides a soliton with nontrivial topology, which is given by the solution of the first-order differential equation
\be\label{eq:BPS}
  \partial_x \phi = \sqrt{2} W'\, .
\ee
Eq.~\eqref{eq:BPS}, which is also called the BPS equation, implies the field equation, since
\be
  \partial_x^2 \phi = \partial_x (\sqrt{2}\, W') = \sqrt{2}\, W'' \,\frac{\partial \phi}{\partial x} = \sqrt{2}\, W''\, \sqrt{2}\, W' = 2 W' W''\, ,
\ee
which is exactly the field equation \eqref{eq:FieldEquation}:
\be
  \partial_x^2 \phi = V' = 2 W' W''\, .
\ee

Now let us come back to the discussion of the soliton solutions to the Gross-Pitaevskii equation. For the repulsive interaction, i.e. $g>0$, the energy for the Gross-Pitaevskii equation is
\be\label{eq:DarkSolEnergy}
  E = \int_{-\infty}^\infty dx \, \left[\frac{\hbar^2}{2m} \bigg|\frac{d\Psi}{dx} \bigg|^2 + \frac{g}{2} \left(|\Psi|^2 - n \right)^2 \right]\, ,
\ee
where
\be
  \Psi = \sqrt{n} \, f\, \textrm{exp} \left[- \frac{i \mu t}{\hbar} \right]
\ee
with the chemical potential $\mu$, and $f$ is in general a complex function
\be
  f = f_1 + i f_2\, .
\ee
We choose $f_2 = \frac{v}{c}$, and define
\be
  \phi \equiv \frac{\hbar}{\sqrt{m}} \Psi\, ,
\ee
then the energy becomes
\be
  E = \int_{-\infty}^\infty dx\, \left[\frac{1}{2} \bigg| \frac{d \phi}{dx} \bigg|^2 + \frac{g}{2} \left(\frac{m}{\hbar^2} |\phi|^2 - n \right)^2 \right]\, .\label{eq:GPenergy}
\ee
Similar to what we discussed before, the BPS equation for the energy functional given by Eq.~\eqref{eq:GPenergy} can be written as follows:
\be
  \frac{d \phi}{dx} = \sqrt{g} \left(n - \frac{m}{\hbar^2} |\phi|^2 \right) \quad \textrm{or} \quad \frac{d \phi}{dx} = \sqrt{g} \left(\frac{m}{\hbar^2} |\phi|^2 - n \right)\, ,
\ee
but the imaginary part of $\phi$ should be constant, in order that the energy functional has the expression of Eq.~\eqref{eq:BPSprocess}. The solutions to these two equations only differ by a minus sign. Let us consider the first equation, which is equivalent to
\begin{align}
  \frac{\hbar}{\sqrt{m}} \frac{df}{dx} & = \sqrt{gn} (1 - |f|^2) \nonumber\\
  \Rightarrow\quad \frac{\hbar}{\sqrt{m}} \frac{df_1}{dx} & = \sqrt{gn} (1 - \frac{v^2}{c^2} - f_1\,^2)\, ,\quad \frac{\hbar}{\sqrt{m}} \frac{i\,df_2}{dx} = 0\, .\label{eq:SolEqBeforeBoost}
\end{align}
For $v=0$ the equations above simplify to
\be
  \sqrt{2} \xi\, \frac{df_1}{dx} = 1 - f_1\,^2\, ,\quad f_2 = 0\, ,
\ee
where $\xi \equiv \hbar / \sqrt{2 m g n}$ is the healing length. The solution to these equations is the dark soliton:
\be
  \Psi (x) = \sqrt{n} \, \textrm{tanh}\, \left[\frac{x}{\sqrt{2} \xi} \right]\, .
\ee

\subsection*{Grey Soliton}
If we perform a Galilean boost to the first one of Eqs.~\eqref{eq:SolEqBeforeBoost} using the method described in Ref.~\cite{Pad}, it becomes
\be
  \sqrt{2} \xi\, \frac{df_1}{dx'} = 1 - \frac{v^2}{c^2} - f_1\,^2\, ,
\ee
where $x' \equiv x - v t$. This new equation is exactly the same as  Eq.~(5.55) in Ref.~\cite{PitaevskiiBook} for an arbitrary constant $v$, and the solution to this equation is
\be
  \Psi (x - v t) = \sqrt{n} \left(i \frac{v}{c} + \sqrt{1 - \frac{v^2}{c^2}}\, \textrm{tanh} \left[\frac{x - v t}{\sqrt{2} \xi} \, \sqrt{1 - \frac{v^2}{c^2}} \right] \right)\, ,
\ee
which includes both the dark soliton solution ($v=0$) and the grey soliton solution ($v \neq 0$).

\subsection*{Bright Soliton}

When the interaction is attractive, i.e. $g<0$, there is another kind of soliton solution to the Gross-Pitaevskii equation, which is called the bright soliton and has the form
\be
  \Psi(x) = \Psi(0) \frac{1}{\textrm{cosh} (x / \sqrt{2} \xi)}\, ,
\ee
where $n_0 = |\Psi(0)|^2$ is the central density, and $\xi \equiv \hbar / \sqrt{2 m |g| n_0}$.

\subsection*{Vortex Line}\label{VortexLine}
The Gross-Pitaevskii equation has another string-like solution called the vortex line. It can be viewed as the Nielsen-Olsen vortex line solution in the Abelian Higgs model in the limit of vanishing gauge field. In this subsection, we follow Ref.~\cite{PitaevskiiBook} to review this kind of solution.

To see the vortex line solution, we start with the Gross-Pitaevskii equation \eqref{eq:GPeq}. Plugging the ansatz
\be
  \phi (\bold{r}, t) = \phi (\bold{r})\, \textrm{exp} \left(-\frac{i \mu t}{\hbar} \right)
\ee
into Eq.~\eqref{eq:GPeq}, where $\mu$ is the chemical potential, we obtain
\be
  \left(-\frac{\hbar^2 \nabla^2}{2m} - \mu + g |\phi (\bold{r})|^2 \right) \phi(\bold{r}) = 0\, .
\ee
For a string-like solution, we can introduce the cylindrical coordinates $(r,\, \varphi,\, z)$ and further parametrize $\phi$ as
\be
  \phi = \sqrt{n} \, f(\eta) \, e^{i s \varphi}\, ,
\label{ansatz}
\ee
where $\eta = r / \xi$ with $\xi \equiv \hbar / \sqrt{2 m |g| n}$, and $s$ is an integer characterizing the angular momentum carried by a vortex line. With this parametrization, one obtains the equation for $f(\eta)$:
\be
  \frac{1}{\eta} \frac{d}{d\eta} \left(\eta \frac{df}{d\eta} \right) + \left(1 - \frac{s^2}{\eta^2} \right) f - f^3 = 0\, ,
\ee
and the boundary conditions are
\begin{align}
  f \to 1\, , & \quad \textrm{when } \eta \to \infty\, ;\nonumber\\
  f \sim \eta^{|s|}\, , & \quad \textrm{when } \eta \to 0\, .
\label{VortexBC}
\end{align}
The equation above can be solved numerically for a given value of $s$. Once the solution $f(\eta)$ is obtained, the energy of this configuration is
\be
  E = \frac{L \pi \hbar^2 n}{m} \int_0^{R / \xi} \eta d\eta \left[\left(\frac{df}{d\eta}\right)^2 + \frac{s^2}{\eta^2} f^2 + \frac{1}{2} \left(f^2 - 1\right)^2 \right]\, ,
\ee
where $L$ and $R$ are the effective length of the vortex line and the radius of the system respectively.

\subsection*{Vortex Ring}
Similar to the vortex line solution discussed in the previous subsection, there is also the vortex ring solution, which does not have two endpoints, instead it is a closed string-like solution. In contrast to the vortex line solution, the vortex ring cannot be at rest. Moreover, as discussed in Ref.~\cite{PitaevskiiBook}, the radius of the vortex ring can be either much larger than the healing length $\xi$ or comparable to the healing length $\xi$. Two parallel vortex rings with opposite circulation can also form a vortex pair, which has been studied in Ref.~\cite{Gubser} using the boson/vortex duality.

\section{Some Details in the Boson/Vortex Duality Map}\label{app:duality}

In this appendix, we present some details of the duality map in the presence of boundaries. Let us start with the Gross-Pitaevskii theory after the parametrization \eqref{eq:Param} given by Eq.~\eqref{eq:LGPfeta}:
\begin{displaymath}
  \mathcal{L} = i p \dot{p} - p^2 \dot{\eta} - \frac{p^2}{2m} (\nabla \eta)^2 - \frac{(\nabla p)^2}{2m} - \frac{g}{2} (p^2 - p_0^2)^2\, ,
\end{displaymath}
where the first term is a total derivative that can be dropped. In the following we analyze the expression above term by term.

\begin{itemize}
\item $-p^2 \dot{\eta}$:

As we analyzed in Section~\ref{sec:duality}, near the soliton plane $p$ changes sign from one side to the other side, while $\eta$ does not have a $\pi$-jump, or in other words, we remove the $\pi$-jump of the phase and allow $p$ to change sign, and now the phase $\eta$ behaves smoothly. Hence, when we consider the limit that the healing length goes to zero, the smooth functions such as $\dot{\eta}$ and $(\nabla \eta)^2$ will take their values at $z=z_0$, where $z_0$ is the longitudinal position of the dark soliton, i.e., the functions of the phase become $z$-independent in this limit. Therefore,
\begin{align}
  - \int d^4 x \, (p_0 + \widetilde{p})^2 \dot{\eta} & = - \left(\int_{z_0 - \ell / 2}^{z_0 + \ell / 2} dz\, p_0^2 \right) \left(\int d^3 x\, \dot{\eta} \right) - \left(\int_{z_0 - \ell / 2}^{z_0 + \ell / 2} dz \right) \left(\int d^3 x\, \widetilde{p}^2 \dot{\eta} \right) \nonumber\\
  {} & \quad - 2 \left(\int_{z_0 - \ell / 2}^{z_0 + \ell / 2} dz\, p_0 \right) \left(\int d^3 x\, \widetilde{p} \dot{\eta} \right)\, ,
\end{align}
where the last term vanishes due to $p_0 (-x) = - p_0 (x)$, and for the first term
\begin{align}
  \int_{z_0 - \ell / 2}^{z_0 + \ell / 2} dz\, p_0^2 & = n \int_{z_0 - \ell / 2}^{z_0 + \ell / 2} dz\, \left[\textrm{tanh} \left(\frac{z - z_0}{\sqrt{2} \ell} \right) \right]^2 \nonumber\\
  {} & = n \int_{- \ell / 2}^{\ell / 2} dz\, \left[\textrm{tanh} \left(\frac{z}{\sqrt{2} \ell} \right) \right]^2 \nonumber\\
  {} & = n \ell \int_{- 1 / 2}^{1 / 2} d\widetilde{z}\, \left[\textrm{tanh} \left(\frac{\widetilde{z}}{\sqrt{2}} \right) \right]^2 \nonumber\\
  {} & = n \ell \left(1 - 2 \sqrt{2} \, \textrm{tanh} (1 / 2 \sqrt{2}) \right) \nonumber\\
  {} & \equiv n \widetilde{\ell}\, .
  \label{scale}
\end{align}
We have defined $\widetilde{z} \equiv z / \ell$, and in the last step we have defined another length scale of the order of the healing length. Therefore,
\be
  - \int d^4 x \, p_0^2 \dot{\eta} = - n \widetilde{\ell} \int d^3 x\, \dot{\eta}\, ,
\ee
which is a total derivative, hence can also be dropped. What remains is
\be
  - \int d^4 x \, (p_0 + \widetilde{p})^2 \dot{\eta} = - \ell \int d^3 x\, \widetilde{p}^2 \dot{\eta}\, .
\ee

\item $- \frac{p^2}{2m} (\nabla \eta)^2$:

Like in the previous case, the smooth function $(\nabla \eta)^2$ becomes $z$-independent in the small region around the dark soliton plane. Hence,
\begin{align}
  - \int d^4 x\,  \frac{(p_0 + \widetilde{p})^2}{2m} (\nabla \eta)^2 & = - \frac{1}{2m} \left(\int_{z_0 - \ell / 2}^{z_0 + \ell / 2} dz\, p_0^2 \right) \left(\int d^3 x\, (\widetilde{\nabla} \eta)^2 \right) \nonumber\\
  {} & \quad - \frac{1}{2m} \left(\int_{z_0 - \ell / 2}^{z_0 + \ell / 2} dz \right) \left(\int d^3 x\, \widetilde{p}^2 (\widetilde{\nabla}  \eta)^2 \right) \nonumber\\
  {} & \quad - \frac{1}{m} \left(\int_{z_0 - \ell / 2}^{z_0 + \ell / 2} dz\, p_0 \right) \left(\int d^3 x\, \widetilde{p} (\widetilde{\nabla}  \eta)^2 \right)\, ,
\end{align}
where $\widetilde{\nabla}$ is the gradient operator on the coordinates $(x,\, y)$. Similar to the previous case, we obtain
\begin{align}
  - \int d^4 x\,  \frac{(p_0 + \widetilde{p} )^2}{2m} (\nabla \eta)^2 & = - \frac{n \widetilde{\ell}}{2m} \int d^3 x\, (\widetilde{\nabla} \eta)^2 - \frac{\ell}{2m} \int d^3 x\, \widetilde{p}^2 (\widetilde{\nabla}  \eta)^2 \nonumber\\
  {} & = - \frac{\ell}{2m} \int d^3 x\, \left(\widetilde{p}^2 + \frac{n \widetilde{\ell}}{\ell} \right)(\widetilde{\nabla} \eta)^2\, .
\end{align}

\item $- \frac{(\nabla p)^2}{2m}$:

\begin{align}
  - \int d^4 x\, \frac{(\nabla p)^2}{2m} & = -\frac{1}{2m} \int_{z_0 - \ell/2}^{z_0 + \ell/2} dz \int d^3 x\, \left[ \left(\frac{\partial \widetilde{p}}{\partial x} \right)^2 + \left(\frac{\partial \widetilde{p}}{\partial y} \right)^2 + \left(\frac{\partial p_0}{\partial z} \right)^2\right] \nonumber\\
  {} & = -\frac{1}{2m} \left(\int_{z_0 - \ell/2}^{z_0 + \ell/2} dz \right) \int d^3 x\, \left[ \left(\frac{\partial \widetilde{p}}{\partial x} \right)^2 + \left(\frac{\partial \widetilde{p}}{\partial y} \right)^2 \right] \nonumber\\
  {} & \quad -\frac{1}{2m} \left[\int_{z_0 - \ell/2}^{z_0 + \ell/2} dz \left(\frac{\partial p_0}{\partial z} \right)^2\right] \left(\int d^3 x \right)\, ,
\end{align}
where the second line in the expression above contributes a constant, which can be dropped from the action, and the first line gives
\be
  -\frac{\ell}{2m} \int d^3 x\, \left( \widetilde{\nabla} \widetilde{p} \right)^2\, .
\ee

\item $- \frac{g}{2} (p^2 - p_c^2)^2$:

\be
  (p^2 - p_c^2)^2 = \left((p_0 + \widetilde{p})^2 - p_c^2\right)^2 = p_0^4 + 4 p_0^3 \widetilde{p} + 6 p_0^2 \widetilde{p}^2 + 4 p_0 \widetilde{p}^3 + \widetilde{p}^4 - 2 p_0^2 p_c^2 - 4 p_0 \widetilde{p} p_c^2 - 2 \widetilde{p}^2 p_c^2 + p_c^4 \, .
\ee
After neglecting the terms that have odd powers in $p_0$, we obtain the relevant terms
\be
  p_0^4 + 6 p_0^2 \widetilde{p}^2 + \widetilde{p}^4 - 2 p_0^2 p_c^2 - 2 \widetilde{p}^2 p_c^2 + p_c^4\, ,
\ee
where the terms independent of $\widetilde{p}$ contribute only constants after the integration over spacetime, which can be dropped from the action. The remaining terms are
\be
  \widetilde{p}^4 - 2 \widetilde{p}^2 (p_c^2 - 3 p_0^2)\, .
\ee

Hence,
\begin{align}
  {} & -\frac{g}{2} \int d^4 x\, (p^2 - p_0^2)^2 \nonumber\\
  = & \, -\frac{g}{2} \left(\int_{z_0 - \ell/2}^{z_0 + \ell/2} dz \right) \left(\int d^3 x\, \widetilde{p}^4 \right) + g \left(\int_{z_0 - \ell/2}^{z_0 + \ell/2} dz\, (p_c^2 - 3 p_0^2) \right) \left(\int d^3 x\, \widetilde{p}^2 \right) \nonumber\\
  = & \, -\frac{g \ell}{2} \int d^3 x\, \widetilde{p}^4 + g (p_c^2 \ell - 3 n \widetilde{\ell}) \int d^3 x\, \widetilde{p}^2 \nonumber\\
  = & \, -\frac{g \ell}{2} \int d^3 x\, \left(\widetilde{p}^2 - n \left(1 - \frac{3 \widetilde{\ell}}{\ell} \right) \right)^2 + \frac{g\ell}{2} \int d^3 x\, \left(n - \frac{3 n \widetilde{\ell}}{\ell} \right)^2\, ,
\end{align}
where we used $p_c = \sqrt{n}$, and the second term above is a constant, that can be dropped from the action. What remains after the integration is
\be
  -\frac{g \ell}{2} \int d^3 x\, \left(\widetilde{p}^2 - n \left(1 - \frac{3 \widetilde{\ell}}{\ell} \right) \right)^2\, ,
\ee
where $\widetilde{\ell} \equiv \ell \left(1 - 2 \sqrt{2} \, \textrm{tanh} (1 / 2 \sqrt{2}) \right)$, hence $(1 - 3 \widetilde{\ell} / \ell)$ is a positive constant. We can define
\be
  \widetilde{p}_c \equiv \sqrt{n \left(1 - \frac{3 \widetilde{\ell}}{\ell} \right)}\, .
\ee

\end{itemize}

Combining all the terms together, we obtain the action around a soliton plane
\begin{align}
  {} & \int d^4 x\, \mathcal{L} \nonumber\\
  = & - \ell \int d^3 x\, \widetilde{p}^2 \dot{\eta} - \frac{\ell}{2m} \int d^3 x\, \left(\widetilde{p}^2 + \frac{n \widetilde{\ell}}{\ell} \right)(\widetilde{\nabla} \eta)^2 - \frac{\ell}{2m} \int d^3 x\, \left( \widetilde{\nabla} \widetilde{p} \right)^2 - \frac{g \ell}{2} \int d^3 x\, \left(\widetilde{p}^2 - \widetilde{p}_c^2 \right)^2 \nonumber\\
  = & \, \ell \int d^3 x\, \left[- \widetilde{p}^2 \dot{\eta} - \frac{1}{2m} \left(\widetilde{p}^2 + C \right)(\widetilde{\nabla} \eta)^2 - \frac{1}{2m} \left( \widetilde{\nabla} \widetilde{p} \right)^2 - \frac{g}{2} \left(\widetilde{p}^2 - \widetilde{p}_c^2 \right)^2  \right] \nonumber\\
  = & \, \ell \int d^3 x\, \left[- \widetilde{\rho} \dot{\eta} - \frac{1}{2m} \left(\widetilde{\rho} + C \right)(\widetilde{\nabla} \eta)^2 - \frac{1}{8m \widetilde{\rho}} \left( \widetilde{\nabla} \widetilde{\rho} \right)^2 - \frac{g}{2} \left(\widetilde{\rho} - \widetilde{\rho}_0 \right)^2 \right]\, ,
\end{align}
where $C \equiv n \widetilde{\ell} / \ell$ is a constant. In the last line we rewrite the theory in the variable $\widetilde{\rho} = \sqrt{\widetilde{p}}$. This action is very similar to the 3D part in the action \eqref{eq:3D4DAction} by restricting the Lagrangian \eqref{eq:LGP} on a 3D space. The only difference is an additional term $- (C / 2m) (\widetilde{\nabla} \eta)^2$, but it does not affect the (2+1)D duality. The reason is following. In the duality map, we will introduce an auxiliary field $f^a = (\rho,\, f^{\hat{a}})$ with $\hat{a} \in \{x,\, y\}$, and for $\widetilde{\rho}\,' \equiv \widetilde{\rho} + C$:
\be
  - \widetilde{\rho} \dot{\eta} - \frac{\widetilde{\rho}\,'}{2m} (\widetilde{\nabla} \eta)^2 + \frac{m}{2\widetilde{\rho}\,'} \left(f_{\hat{a}} - \frac{\widetilde{\rho}\,'}{m} \widetilde{\nabla}_{\hat{a}} \eta \right)^2 = - \widetilde{\rho} \dot{\eta} + \frac{m}{2 \widetilde{\rho}\,' } f^{\hat{a}} f_{\hat{a}} - f^{\hat{a}} \partial_{\hat{a}} \eta = \frac{m}{2 \widetilde{\rho}\,' } f^{\hat{a}} f_{\hat{a}} - f^a \partial_a \eta\, ,
\ee
where in the path integral
\be
  \int \mathcal{D} f^{\hat{a}}\, \textrm{exp} \left(i \ell \int d^3 x\, \frac{m}{2 \widetilde{\rho}\,' } f^{\hat{a}} f_{\hat{a}} \right) = 1\, .
\ee
Hence, $\widetilde{\rho}\,'$ or consequently the constant $C$ does not show up in the action after the duality map.

\section{Tachyon Potential}\label{app:TachyonPot}

In this appendix we discuss how to compute the tachyon potential $V(T)$ appearing in the effective action \eqref{eq:tachyonAction} from string field theory \cite{WittenTachyon-1, WittenTachyon-2, Shatashvili-3, KutasovMarino,Cornalba:2000ad,Okuyama:2000ch}.

Consider the string action defined on the unit disk $\Sigma$ given by
\be
S=S_{0}+S'\, ,
\ee
where $S_{0}$ is a bulk action and $S'$ is a boundary term. In particular,
\be
S_{0}=\frac{1}{4\pi\alpha'}\int_{\Sigma}\left(g_{\mu\nu}\partial_{a}X^{\mu}\partial^{a}X^{\nu}-2\pi i\alpha'B_{ij}\epsilon^{ab}\partial_{a}X^{i}\partial_{b}X^{j}\right)\, ,
\ee
where $\mu,\nu=0,1,\ldots,D$ and $i,j=0,1,\ldots,p\leq D$. For the time being we can ignore the directions $\mu=p+1,\ldots,D$. For a constant $B$-field,
\be
-\frac{i}{2}\int_{\Sigma}B_{ij}\epsilon^{ab}\partial_{a}X^{i}\partial_{b}X^{j}=-\frac{i}{2}\int_{\partial\Sigma}B_{ij}X^{i}\partial_{t}X^{j}\, ,
\ee
where $\partial_{t}$ is a tangential derivative along the boundary $\partial\varSigma$. The boundary conditions determined by the equations of motion are
\be
\left.g_{ij}\partial_{n}X^{j}+2\pi i\alpha'B_{ij}\partial_{t}X^{j}\right|_{\partial\varSigma}=0\, ,
\ee
where $\partial_{n}$ is a normal derivative to $\partial\Sigma$. For $B=0$ these are Neumann boundary conditions, corresponding to
open strings, and that is why we can refer to $g_{ij}$ as the closed
string metric. When $B$ has rank $r=p$ and $B\rightarrow\infty$,
or equivalently $g_{ij}\rightarrow0$, along the spatial directions
of the brane, the boundary conditions become Dirichlet, i.e. $\left.\partial_{t}X^{j}\right|_{\partial\varSigma}=0$.
Therefore, the physical picture is that for $B=0$ the ends of the
open string are free to move, and the Polyakov action describes
the space-filling $D_p$-brane.

In the following discussions, it is more convenient to use the two open string parameters
\begin{eqnarray*}
G^{-1} & \equiv & \left(\frac{1}{g+2\pi i\alpha'B}\right)_{S}\, ,\\
\Theta & \equiv & \left(\frac{1}{g+2\pi i\alpha'B}\right)_{A}\, ,
\end{eqnarray*}
which are symmetric and antisymmetric respectively. In Ref. \cite{WittenTachyon-1,WittenTachyon-2} it has been argued
that the open strings are described by the boundary term $S'$, which
has the form
\be
S'
=\intop_{0}^{2\pi}\frac{d\sigma}{2\pi}\mathcal{V}\, ,
\ee
where $\sigma$ is a parameter on the border $\partial \Sigma$, and $\mathcal{V}$ is a general boundary perturbation that can be parametrized by couplings $\lambda^{i}$:
\be
\mathcal{V}=\lambda^{i}\mathcal{V}_{i}\, .
\ee
Defining the ghost number-one operator $\mathcal{O}=c\mathcal{V}$,
the spacetime string field theory action $S$ is defined by
\be
\frac{\partial S}{\partial\lambda^{i}}=\frac{1}{2}\intop_{0}^{2\pi}\frac{d\sigma}{2\pi}\intop_{0}^{2\pi}\frac{d\sigma'}{2\pi}\left\langle \mathcal{O}_{i}\left(\sigma\right)\left\{ \mathcal{Q}_{B},\mathcal{O}\left(\sigma'\right)\right\} \right\rangle _{\lambda}\, ,
\ee
where $\mathcal{Q}_{B}$ is the BRST charge. For the tachyon field $\mathcal{O}=cT\left(X\right)$,
one can find 
\be
\left\{ \mathcal{Q}_{B},cT\left(X\right)\right\} =c\partial_{t}c\left(1-\Delta_{T}\right)T\left(X\right)\, ,
\ee
where 
\be
\Delta_{T}=-\alpha'G^{ij}\frac{\partial^{2}}{\partial X^{i}\partial X^{j}}\, .
\ee
The general form for the action satisfying this equation is \cite{Shatashvili-1, Shatashvili-2}:
\be
S=-\beta^{i}\frac{\partial Z}{\partial\lambda^{i}}+Z\, ,
\ee
with $Z$ the partition function and $\beta^{i}$ the beta function for the coupling $\lambda^{i}$. In particular, for the following explicit form of the tachyon profile \cite{Shatashvili-3, Okuyama:2000ch}:
\be
T\left(X\right)=a+\frac{1}{2\alpha'}u_{ij}X^{i}X^{j}\, ,
\ee
one can rewrite the action as
\be
S\left(a,u\right)=\left[tr\left(G^{-1}u\right)-a\frac{\partial}{\partial a}-tr\left(u\frac{\partial}{\partial u}\right)+1\right]Z\left(a,u\right)\, ,
\ee
where 
\be
Z\left(a,u\right)=e^{-a+\gamma tr\left(G^{-1}u\right)}\det\phantom{}^{\frac{1}{2}}\left(\varGamma\left(E_{+}u\right)\varGamma\left(1+E_{-}u\right)\right)\, ,
\ee
with $E_{\pm}=G^{-1}\pm\varTheta$.

There are two equivalent descriptions of the action above. We can describe the $D_p$-brane in a constant $B$-field background by treating the $B$-term as a boundary interaction term. In this approach, the boundary conditions are Neumann and by direct computation one can find
\be
Z\left(a,u\right)=T_{D_p}\int d^{p+1}xe^{-T}\sqrt{\det\left(g+2\pi\alpha'B\right)}\left(1+\ldots\right)\, ,
\ee
where the dots stand for the higher-order terms in $u$ or the higher-derivative terms of $T$. One can therefore reconstruct the following shape for the action 
\be
S=T_{D_p}\int d^{p+1}xe^{-T}\sqrt{\det\left(g+2\pi\alpha'B\right)}\left(1+T+\alpha'G^{ij}\partial_{i}T\partial_{j}T+\ldots\right)\, .
\ee
This form, even if reproducing the standard gauge symmetries and showing the connection to the well-known DBI action, is not particularly helpful in studying the soliton solutions. Alternatively, the action takes a much more convenient form in the large noncommutativity limit studied by Seiberg and Witten \cite{SW}, i.e. $G^{-1}\ll\varTheta$, in which the partition function becomes
\be
\lim_{G\Theta\rightarrow\infty}Z\left(a,u\right)=e^{-a}\det\phantom{}^{\frac{1}{2}}\left(\frac{\pi}{\sin\pi\varTheta u}\right)\, .
\ee
As shown in Refs.~\cite{Cornalba:2000ad,Okuyama:2000ch}, this can be written conveniently as
\be
\int\frac{d^{p}x}{\textrm{Pf}\left(2\pi\theta\right)}\exp_{\star}\left(-T\left(x\right)\right)\, ,
\ee
with $\theta^{ij}=2\pi\alpha'\Theta^{ij}$. It is clear that in this limit the kinetic term is suppressed, and the action is dominated by the potential term:
\be
S=\int\frac{d^{p}x}{\textrm{Pf}\left(2\pi\theta\right)}\left(T\left(x\right)+1\right)\star\exp_{\star}\left(-T\left(x\right)\right)\, .
\ee
From this expression, we can read off the well-known form of the tachyon potential \cite{WittenTachyon-1,WittenTachyon-2, Shatashvili-3, KutasovMarino}:
\be
  V(T) = (T+1)\, e^{-T}\, .
\ee

\bibliographystyle{utphys}
\bibliography{BECstring}

\end{document}